\documentclass[reprint,superscriptaddress,nofootinbib,amsmath,amssymb,aps,floatfix]{revtex4-1}
\pdfoutput=1
\usepackage{graphicx}
\usepackage{dcolumn}
\usepackage{bm}
\usepackage[utf8]{inputenc}
\usepackage{graphics,graphicx}
\usepackage{parskip}
\usepackage{amsmath,amssymb}
\usepackage[normalem]{ulem}
\usepackage{amsfonts}

\usepackage{multirow}
\usepackage{fancyhdr}
\usepackage{xcolor}
\usepackage{placeins}
\usepackage[title]{appendix}
\usepackage{wasysym}
\usepackage{url}
\usepackage{braket}
\usepackage{float}
\usepackage[caption=false]{subfig}

\usepackage{lipsum}
\usepackage{booktabs}
\usepackage{comment}
\usepackage{braket}
\usepackage[colorlinks=true,filecolor=.]{hyperref}
\usepackage{cleveref}
\usepackage{natbib}
\usepackage{setspace}
\usepackage{subfiles} 
\newcommand{\figwidth}{0.725\textwidth}
\begin{document}

\title{Charged ultralong-range Rydberg trimers}

\author{Daniel J. Bosworth} 
 \email{dboswort@physnet.uni-hamburg.de}
\affiliation{%
 Zentrum f\"ur Optische Quantentechnologien, Universit\"at Hamburg,\\ Luruper Chaussee 149, 22761 Hamburg, Germany\\
}%
\affiliation{%
 The Hamburg Centre for Ultrafast Imaging, Universit\"at Hamburg,\\ Luruper Chaussee 149, 22761 Hamburg, Germany\\
}%

\author{Frederic Hummel}%
\email{hummel@pks.mpg.de}
\affiliation{%
 Max Planck Institute for the Physics of Complex Systems,\\
Nöthnitzer Straße 38,
01187 Dresden, Germany
}%

\author{Peter Schmelcher}
\affiliation{%
 Zentrum f\"ur Optische Quantentechnologien, Universit\"at Hamburg,\\ Luruper Chaussee 149, 22761 Hamburg, Germany\\
}%
\affiliation{%
 The Hamburg Centre for Ultrafast Imaging, Universit\"at Hamburg,\\ Luruper Chaussee 149, 22761 Hamburg, Germany\\
}%

\date{\today}

\begin{abstract}
We show that the recently observed class of long-range ion-Rydberg molecules can be divided into two families of states, which are characterised by their unique electronic structures resulting from the ion-induced admixture of quantum defect-split Rydberg $n$P states with different low-field seeking high-$l$ states. We predict that in both cases these diatomic molecular states can bind additional ground state atoms lying within the orbit of the Rydberg electron, thereby forming \textit{charged} ultralong-range Rydberg molecules (ULRM) with binding energies similar to that of conventional non-polar ULRM. To demonstrate this, we consider a Rydberg atom interacting with a single ground state atom and an ion. The additional atom breaks the system's cylindrical symmetry, which leads to mixing between states that would otherwise be decoupled. The electronic structure is obtained using exact diagonalisation over a finite basis and the vibrational structure is determined using the Multi-Configuration Time-Dependent Hartree method. Due to the lobe-like structure of the electronic density, bound trimers with both linear and nonlinear geometrical configurations of the three nuclei are possible. The predicted trimer binding energies and excitation series are distinct enough from those of the ion-Rydberg dimer to be observed using current experimental techniques.
\end{abstract}

\maketitle

\section{Introduction}
Hybrid atom-ion systems serve as a test-bed for fundamental quantum physics research~\cite{Zipkes2011Hybrid,Tomza2019Cold,Lous2022Chapter}, enabling studies of cold collisions and chemistry \cite{Ratschbacher2012Controlling,Hall2012Millikelvin,Harter2012Single,Meir2016Dynamics,Perez-Rios2021Cold,Oghittu2022Quantumlimited} such as the formation of cold molecular ions~\cite{Cote2002Mesoscopic,Schurer2017Unraveling,Bosworth2021Spectral}. They also provide a platform for precision measurements~\cite{Schmid2010Dynamics,Zipkes2010Trapped,Veit2021Pulsed} and quantum simulation~\cite{Gerritsma2012Bosonic,Schurer2016Impact}. Recent milestones in this field include the first reports of $s$-wave atom-ion collisions~\cite{Feldker2020Buffer} and the observation of atom-ion Feshbach resonances~\cite{Weckesser2021Observation}.\\
Over the last two decades, there has also been growing interest in combining ions with Rydberg atoms in order to engineer atom-ion interactions~\cite{Secker2016Controlled,Ewald2019Observation} and control cold collisions and charge hopping~\cite{Secker2017Trapped,Schmid2018Rydberg,Cote2000Classical,Dieterle2021Transport}. Additionally, an ion-induced Rydberg blockade effect has been established~\cite{Engel2018Observation} and Rydberg atoms have been used to realise hybrid atom-ion systems in the quantum regime without the need for an ion trap~\cite{Kleinbach2018Ionic}. Recent theoretical works predicted bound molecular states between ions and Rydberg states of Rb and Cs~\cite{Duspayev2021Longrange,Deiss2021LongRange} and their existence and vibrational dynamics were observed shortly after~\cite{Zuber2022Observation,Zou2022Observation}. These ion-Rydberg molecules have bond lengths and energies ranging from nm to $\mu$m and MHz to GHz, respectively.\\ 
Such extreme bonding lengths and energies are seen in another exotic type of Rydberg molecule formed between a Rydberg atom and one or more ground state atoms which become bound due to attractive electron scattering. These are known as ultralong-range Rydberg molecules (ULRM)~\cite{Eiles2019Trilobites,Fey2020Ultralongrange}. The importance of electron scattering for describing Rydberg atoms in atomic gases originated with Fermi~\cite{Fermi1934Sopra}. Fermi's model was later applied within the context of ultracold atomic gases, leading to the prediction of ULRM in 2000~\cite{Greene2000Creation}. These molecules were first observed in 2009~\cite{Bendkowsky2009Observation} and over the past decade they have been used for probing spatial correlations in ultracold atomic gases~\cite{Manthey2015Dynamically,Whalen2019Probing,Whalen2019Formation}, studying low-energy electron-atom collisions~\cite{Anderson2014Photoassociation,Sassmannshausen2015Experimental,Bottcher2016Observation} and the formation of Rydberg polarons in the high-density regime~\cite{Schmidt2016Mesoscopic,Camargo2018Creation,Sous2020Rydberg,Kleinbach2018Ionic}. Furthermore, the formation of polyatomic ULRM and Rydberg composites has been a topic of major interest~\cite{Bendkowsky2010Rydberg,Gaj2014Molecular,Liu2009UltraLongRange,Fey2016Stretching,Eiles2016Ultracold,Fey2019Effective,Eiles2020Triatomic,Hunter2020Rydberg} as well as the behaviour of ULRM in external electric and magnetic fields~\cite{Lesanovsky2006Ultralongrange,Kurz2014Ultralongrange,Krupp2014Alignment,Gaj2015Hybridization,Niederprum2016Observation,Hummel2019Alignment,Engel2019Precision,Hummel2021Electricfieldinduced}.\\
\begin{figure*}[t]
    \centering
    \includegraphics[width=\figwidth]{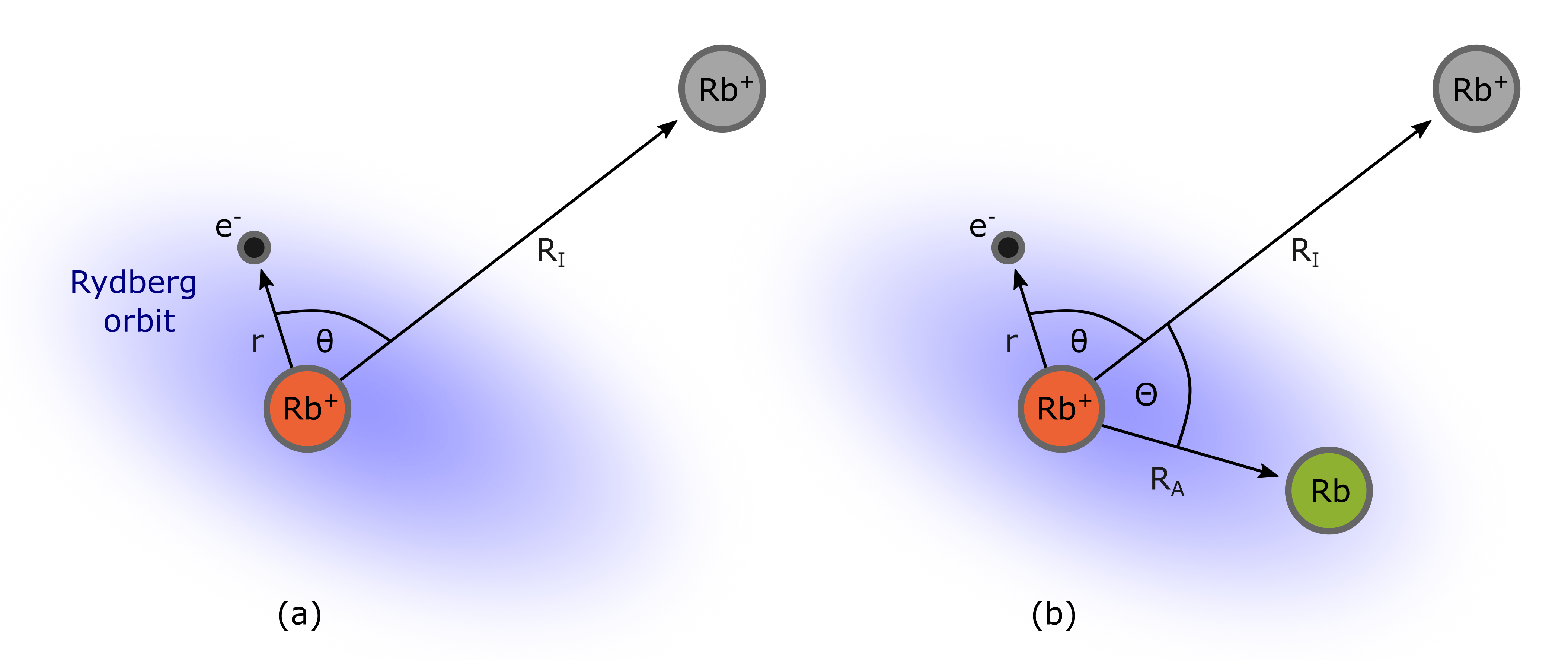}
    \caption{Schematic illustration of the two- and three-body systems. (a) a Rydberg atom (ionic core and valence electron) interacting with an ion. (b) a Rydberg atom interacting with an ion and a ground state atom.}
    \label{fig:schematic}
\end{figure*}
In light of recent developments, combining ULRM with ions is a natural step forward: it constitutes a system for studying Rydberg molecules exposed to \textit{inhomogeneous} electric fields created by the ion and in which Rydberg atoms may become bound to both ions and ground state atoms simultaneously. In contrast to neutral ULRM where the ground state atom is bound within the Rydberg cloud, ions bind with Rydberg atoms far outside the orbit of the Rydberg electron. The ion-Rydberg potential wells are formed due to couplings between neighbouring electronic states, reminiscent of Rydberg macrodimer potential wells~\cite{Hollerith2022Rydberg}.\\
In this work, we begin by examining the electronic structure of a two-body system consisting of an ion and a Rydberg atom. We focus in particular on the electronic density distributions of ion-Rydberg bound states that exist within different potential wells present in the adiabatic potential energy curves (PEC). These reveal patterns of pronounced density maxima that are unique to each potential well due to the ion-induced admixture of different high- and low-$l$ Rydberg states. Despite their differences, we show that the electronic density structures can be grouped into two families of states, which differ primarily in their angular arrangement of the density maxima. We remark that the Rydberg electron's probability density in one of these wells has been discussed previously within the context of the Rydberg atom's flipping electric dipole moment around the well minimum and the electronic oscillations accompanying nuclear dynamics in this well~\cite{Zuber2022Observation,Zou2022Observation}. Building on our considerations of the two-body system's electronic structure, we then introduce an additional ground state atom into the system. We determine adiabatic potential energy surfaces (PES) near the 32P atomic Rydberg state and uncover that their local minima support weakly-bound trimer states with unique excitation spectra for different geometrical arrangements of the three species. The properties of the trimer states are compared against those of the dimer for principal quantum numbers ranging from 17 to 90.\\
This work is organised as follows. Section~\ref{sec:dimer} presents an overview of the current understanding of ion-Rydberg molecules as well as a discussion of their electronic structure. In section~\ref{sec:timer-electronic}, we introduce our three-body system and analyse the corresponding adiabatic PES. Section~\ref{sec:timer-vib} discusses the resulting vibrational states supported by these surfaces and compares their properties to those of ion-Rydberg dimers, including possible experimental aspects. Our conclusions are provided in section~\ref{sec:summary}.\\ 

\section{Electronic structure of ion-Rydberg dimers}~\label{sec:dimer}
This section first provides some background information on ion-Rydberg molecules. We then present results for the electronic structure of different molecular states which motivate the discussion of the three-body system in sections~\ref{sec:timer-electronic} and~\ref{sec:timer-vib}.
\subsection{Squid and snow angel states}\label{ssec:dimer}
\begin{figure*}[t]
    \centering
    \includegraphics[width = \figwidth]{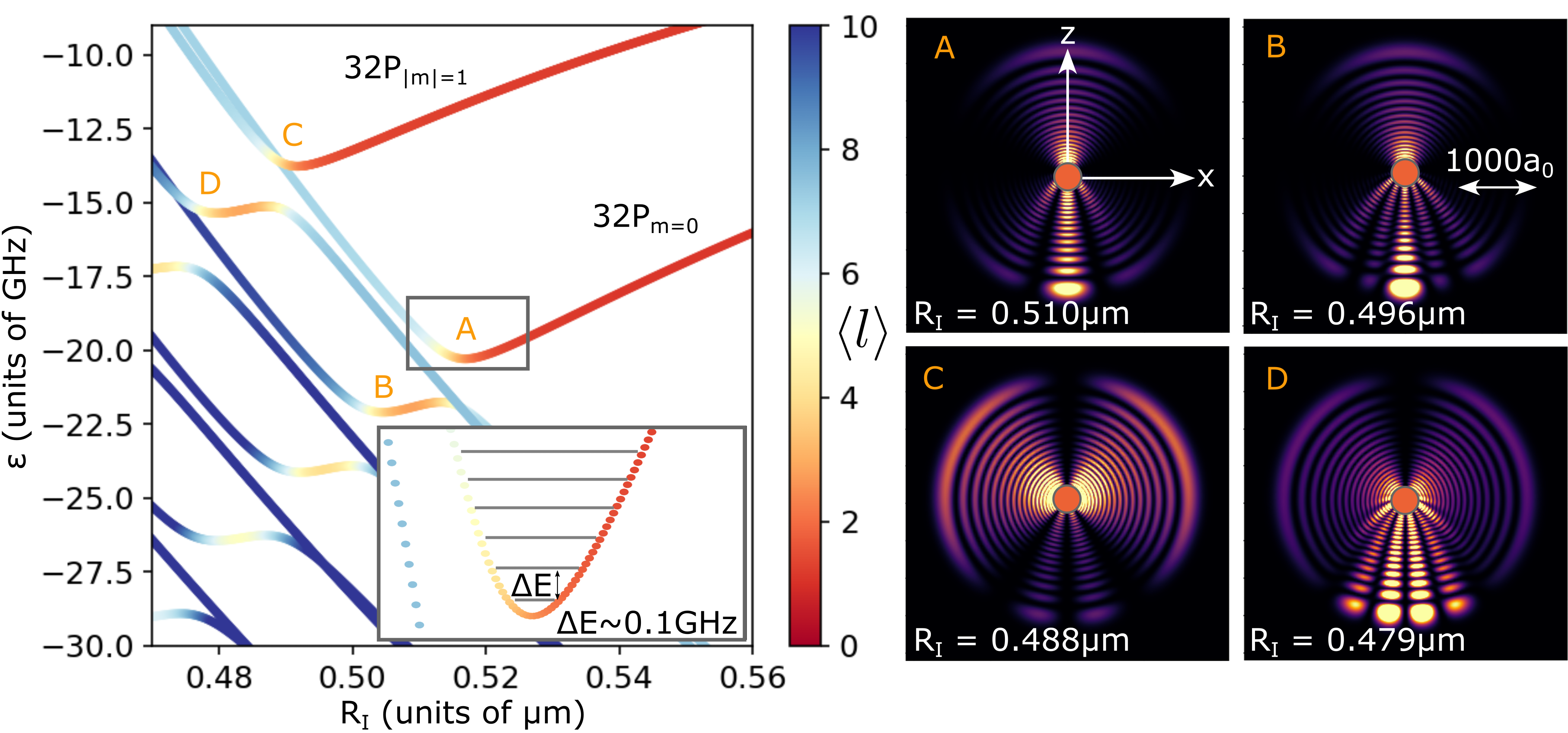}
    \caption[Ion-Rydberg dimer potential energy curves and electronic density structure.]{Adiabatic potential energy curves and electronic density of the ion-Rydberg dimer system. Left: adiabatic PEC near the 32P atomic Rydberg state. The colourbar indicates the $l$-character $\braket{l}$ of the corresponding electronic states. Energies are given relative to the field-free 32P atomic Rydberg state. Inset: close-up of well A showing the energies of the first few vibrational states. Right: the electronic density $|\psi(r,\theta,\phi)|^2$ in the $y=0$ plane at the four minima A-D shown in (a). Plots are normalised such that $\int r^2 \sin\theta |\psi(r,\theta,\phi)|^2dr d\theta d\phi = 1$. Note that the ion (not shown) lies outside the Rydberg cloud.}
    \label{fig:dimer-pecs}
\end{figure*}
We consider a Rydberg atom in the presence of an ion at internuclear distances in which there is vanishing spatial overlap between the charge distribution of the Rydberg atom and the ion, as illustrated in Fig.~\ref{fig:schematic}~(a). This holds for internuclear distances greater than the so-called LeRoy radius $R_{LR}$~\cite{LeRoy1974LongRange}, which for this system is approximately defined by the diameter of the Rydberg cloud.\\
The system's electronic structure is described by the Hamiltonian $H_e = H_0 + V_{ei}(R_I)$~\footnote{Unless stated otherwise, atomic units are assumed throughout.}. $H_0$ describes the Rydberg electron interacting with the positively-charged Rydberg core and $V_{ei}(R_I)$ is the net multipole interaction between the ion and Rydberg atom
for a given internuclear separation $R_I$~\cite{Duspayev2021Longrange,Deiss2021LongRange}
\begin{equation}\label{eq:ion-ryd-int}
    V_{ei}(R_I) = -\sum_{\lambda=1}^{\infty} \sqrt{\frac{4\pi}{2\lambda + 1}}\frac{r^\lambda}{R_I^{\lambda+1}} Y_{\lambda}^{0}(\theta,\phi).
\end{equation}
We choose our coordinate system such that the Rydberg core is located at the origin and the ion-Rydberg internuclear axis lies along the $z$-axis. The position of the Rydberg electron relative to the Rydberg core is given by $(r,\theta,\phi)$ in spherical coordinates. $Y_{\lambda}^{\mu}$ are the spherical harmonics with angular momentum $\lambda$ and angular momentum projection $\mu$, which should not be confused with the orbital angular momentum quantum numbers defining the Rydberg state, namely $l$ and $m$. Since we have chosen $\mathbf{R}_I=R_I\hat{z}$, $\mu$ is restricted to $\mu=0$. The order of the multipole expansion in~\eqref{eq:ion-ryd-int} is typically truncated at $\lambda=6$, since higher-order terms only provide energy corrections on the sub-MHz level which can be safely neglected here. \\ 
The time-independent Schr\"{o}dinger equation for the electronic Hamiltonian reads $H_e\psi_{\nu}(\mathbf{r};R)=\varepsilon_{\nu}(R)\psi_{\nu}(\mathbf{r};R)$, which depends parametrically on the internuclear separation $R$ whilst $\nu$  labels the separate adiabatic electronic states. The Born-Oppenheimer PEC $\{\varepsilon_{\nu}\}$ obtained from the exact diagonalisation of $H_e$ in a finite Rydberg basis are shown in the region of the Rydberg 32P state in Fig.~\ref{fig:dimer-pecs}. Note that we neglect the fine and hyperfine structures in our analysis since they are not responsible for the primary features and results obtained here.\\
As predicted in~\cite{Duspayev2021Longrange,Deiss2021LongRange}, the ion-Rydberg interaction potential $V_{ei}(R_I)$ couples nearby Rydberg states of different angular momentum $l$ character, leading to a series of potential wells in the vicinity of the Rydberg $p$-state which support bound vibrational states with a spacing on the order of 100~MHz (see inset of Fig.~\ref{fig:dimer-pecs}). The colourbar in Fig.~\ref{fig:dimer-pecs} encodes the $l$-character of the electronic states, which changes in particular around the avoided crossings. In principal, these avoided crossings should introduce non-adiabatic corrections to the Born-Oppenheimer approximation, yet remarkably the vibrational energies calculated using the adiabatic approximation are in excellent agreement with current measurements~\cite{Zuber2022Observation}. Moreover, a recent theoretical study determined that the non-adiabatic decay rate of ion-Rydberg molecules should be far smaller than the radiative decay of the parent Rydberg atom~\cite{Duspayev2022Nonadiabatic}.\\
As a direct result of the admixture of low- and high-$l$ states, the electronic densities of these molecules display interesting lobe-like patterns. These can be seen in Fig.~\ref{fig:dimer-pecs}, which show 2D slices of the Rydberg electron's probability density in the $y=0$ plane for the first four potential wells, marked A to D. The lobes differ in the degree of their azimuthal localisation. Additionally, the number of lobes present in the probability density increases for potential wells lying deeper in the fan of electronic states, thereby forming a series of unique electronic densities. We choose to classify the series of electronic densities formed by states of $m = 0$ character (Fig.~\ref{fig:dimer-pecs} A and B) as `squid states' from their resemblance to a head with several appendages, whilst the wing-like features of the density patterns formed by $|m| = 1$ states (Fig.~\ref{fig:dimer-pecs} C and D) are reminiscent of an angel pattern made in the snow.\\
We emphasise that the potential wells shown near the 32P atomic Rydberg state in Fig.~\ref{fig:dimer-pecs} are a general feature of the ion-Rydberg system and should appear in the adiabatic PEC over a wide range of principal quantum numbers. Indeed, molecular states have already been observed at different principle quantum numbers~\cite{Zuber2022Observation}. Furthermore, a perturbative treatment of the ion-Rydberg interaction given by Eq.~\eqref{eq:ion-ryd-int} yields leading-order energy corrections to the Rydberg $p$-state and quasi-degenerate high-$l$ states of $\varepsilon_{p}\propto-n^7/R^4$ and $\varepsilon_{l>3}\propto \pm n^2/R^2$, respectively. From these results, it is expected that the equilibrium separation and binding energy of the ion-Rydberg molecule should scale as $R_e\propto n^{2.5}$ and $\varepsilon_b\propto n^{-3}$, similar to the scaling laws of Rydberg macrodimer binding potentials~\cite{Hollerith2022Rydberg}.\\

\begin{figure*}[t]
    \centering
    \includegraphics[width = \figwidth]{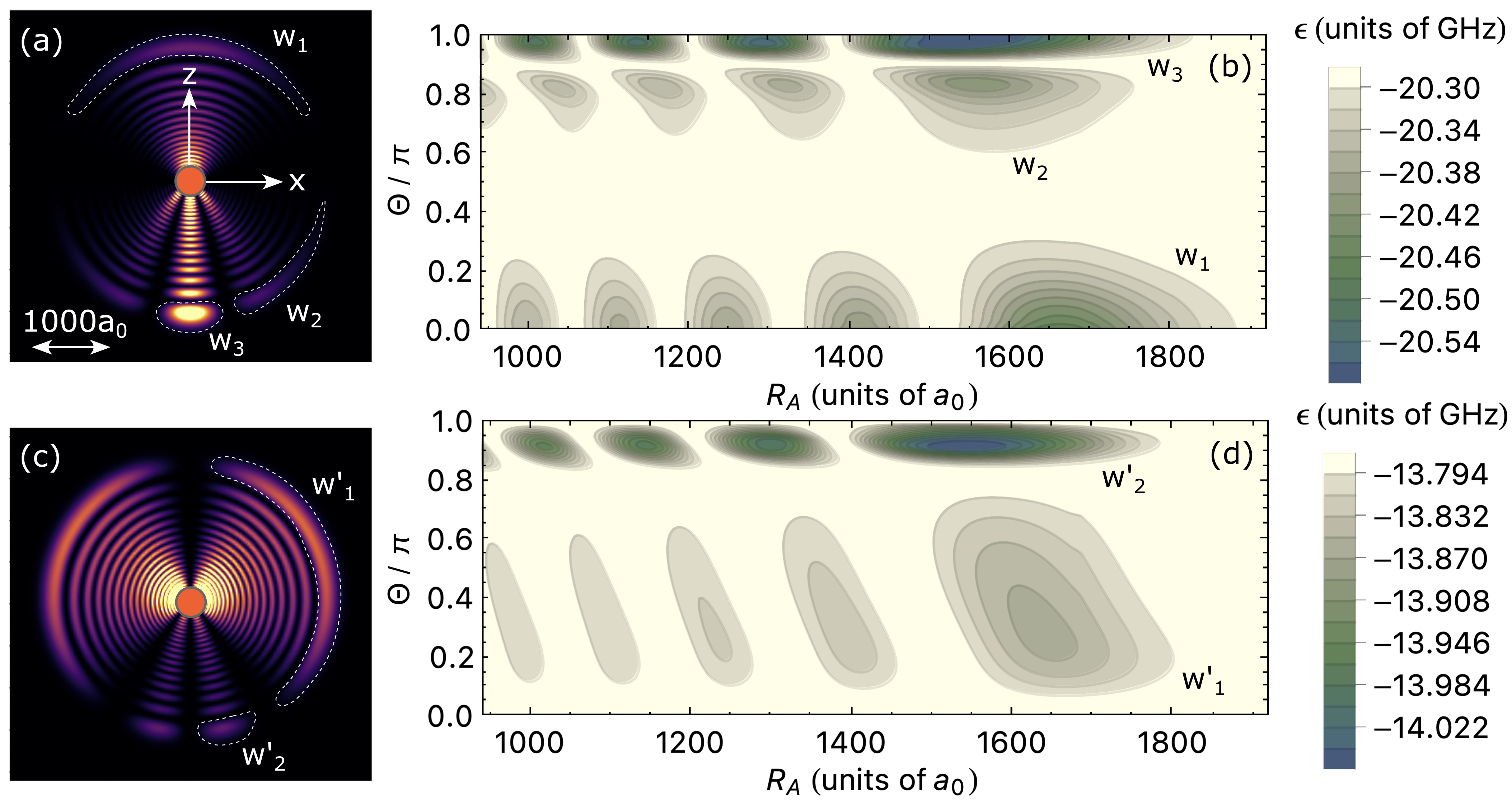}
    \caption[Potential energy surfaces of the three-body system.]{Adiabatic potential energy surfaces of the squid and snow angel states for the trimer system. (a) shows the electronic density of the three-legged squid state in the dimer system. (b) is a 2-D slice taken at $R_I = 0.510~\mu$m of the squid state's PES with major local minima denoted by \{$\text{w}_i$\}. (c) shows the electronic density of the snow angel state in the $y=0$ plane. (d) is a 2-D slice taken at $R_I = 0.488~\mu$m of the snow angel state's PES with major local minima denoted by \{$\text{w}^{\prime}_i$\}. Energies are given relative to the field-free 32P atomic Rydberg state.}
    \label{fig:trimer-pes}
\end{figure*}
\subsection{From the charged dimer to the trimer}
The interaction of a Rydberg atom and a ground state atom is determined by the scattering of the highly-excited Rydberg electron off the ground state atom. The corresponding $s$-wave scattering term is described by the Fermi pseudopotential $V_{ea} = 2\pi a_s[k(R_A)]\delta(\mathbf{r}-\mathbf{R_A})$. This model predicts the appearance of wells in the PEC for regions where the scattering length $a_s[k(R_A)]$ is negative. These wells support weakly-bound vibrational states with binding energies in the range of MHz to GHz as first shown in~\cite{Greene2000Creation}. These bound molecular states are \textit{ultralong-range Rydberg molecules} (ULRM) and examples include the polar trilobite state and the non-polar $s$-state ULRM~\cite{Eiles2019Trilobites}.\\
Motivated by this fact, we now turn to the question of whether the same mechanism could enable ion-Rydberg molecules to bind an additional ground state atom within the orbit of the Rydberg electron, forming a \textit{charged} ULRM trimer. To that end, we consider an additional ground state atom within the Rydberg cloud, as shown by the schematic in Fig.~\ref{fig:schematic} (b). If this binding is possible, the various lobes of high electron density present in the squid and snow angel states would enable the formation of both linear and nonlinear bound configurations of the atoms, akin to the $d$-state angular trimers reported in~\cite{Fey2019Effective}.\\

\section{Electronic structure of the trimer}~\label{sec:timer-electronic}
In this section, we consider the effect of an additional ground state atom on the first squid and snow angel states, whose unperturbed electronic densities are shown in Fig.~\ref{fig:dimer-pecs} A and C, respectively.
\subsection{Setup and interactions}\label{ssec:setup}
With the Rydberg core at the origin of our coordinate system, the ion and ground state atom are located at positions $(R_I,0,0)$ and $(R_A,\Theta,0)$, respectively. A schematic for this three-body system is provided in Fig.~\ref{fig:schematic} (b). Within the Born-Oppenheimer approximation, the electronic structure of the Rb$^+$-Rb$^*$-Rb system is based on the following Hamiltonian
\begin{equation}\label{eq:trimer-eham}
    \begin{split}
    H_e &= H_0 +V_{ei}(R_I)\\
    	 &+ V_{ea}(R_A) + V_{ca}(R_A) + V_{ia}(R_I,R_A, \Theta).
    \end{split}
\end{equation}
$V_{ei}$ describes the electron-ion interaction~\eqref{eq:ion-ryd-int} and $V_{ea}$ represents the $s$-wave scattering between the electron and the ground state atom. $V_{ca}$ and $V_{ia}$ are the interactions between the Rydberg core and the ion with the neutral atom, respectively. These interactions take the form of a charge-induced dipole (polarisation potential) interaction, e.g. $V_{ca} \propto -1/R_A^4$.\\
We obtain the adiabatic PES $\{\varepsilon_{\nu}(R_I,R_A,\Theta)\}$ from the exact diagonalisation of the electronic Hamiltonian~\eqref{eq:trimer-eham} using a finite basis of unperturbed Rydberg states $\{\ket{n,l,m}\}$, where the quantum numbers take their usual meaning. Specifically, our electronic basis includes all states enclosed by the nearest six hydrogenic manifolds centred around the 32P atomic Rydberg state up to a maximum magnetic quantum number of $|m|=4$.\\
Besides introducing additional interactions, the presence of the ground state atom breaks the dimer's cylindrical symmetry, such that $m$ is no longer a good quantum number. This means couplings between states of different $m$-character are no longer prohibited, though generally this will only be relevant where states become near-degenerate. One example of this occurs  close to the minimum of the first snow angel state, labelled as C in Fig.~\ref{fig:dimer-pecs}. In principle, this requires that we enlarge our basis to account for the additional couplings that arise between different $m$-states. \\
\subsection{Adiabatic potential energy surfaces}
Let us analyse in the following the relevant adiabatic PES of the trimer. Figure~\ref{fig:trimer-pes} shows the electronic densities of the squid and snow angel states in the dimer system alongside 2D slices of the corresponding Born-Oppenheimer PES $\varepsilon(R_I,R_A,\Theta)$ for the trimer system at fixed $R_I$. Each lobe in the electronic density leads to a unique local minimum along $\Theta$. Along $R_A$, the surfaces exhibit a series of local minima due to the oscillatory electronic density along each lobe. The deepest minima are found at $R_A\approx 1700~a_0$ and have a depth on the order of 100~MHz, similar to a conventional non-polar ULRM at this principal quantum number.\\ 
For the squid state we focus on three local minima labelled w$_1$, w$_2$ and w$_3$ in Fig.~\ref{fig:trimer-pes} (b). These minima occur at angles $\Theta=0$, $\Theta=0.8\pi$ and $\Theta=\pi$, respectively which map directly to the three unique lobes in the electronic density in Fig.~\ref{fig:trimer-pes} (a). Well w$_3$ is the deepest of these, which is not surprising since the electronic probability density is highly-localised at this position. For the snow angel state, we focus on the two minima labelled w$^{\prime}_1$ and w$^{\prime}_2$. The depths of these wells are similar in magnitude to those of the squid state. Since the electronic probability density vanishes along the $z$-axis, no local minima appear along $\Theta=0$ and $\Theta=\pi$ and as such it should not be possible for the snow angel to support linear trimer configurations.\\
\subsection{Coupling among states of different magnetic quantum number}
\begin{figure}[t]
    \centering
    \includegraphics[width = 0.4\textwidth]{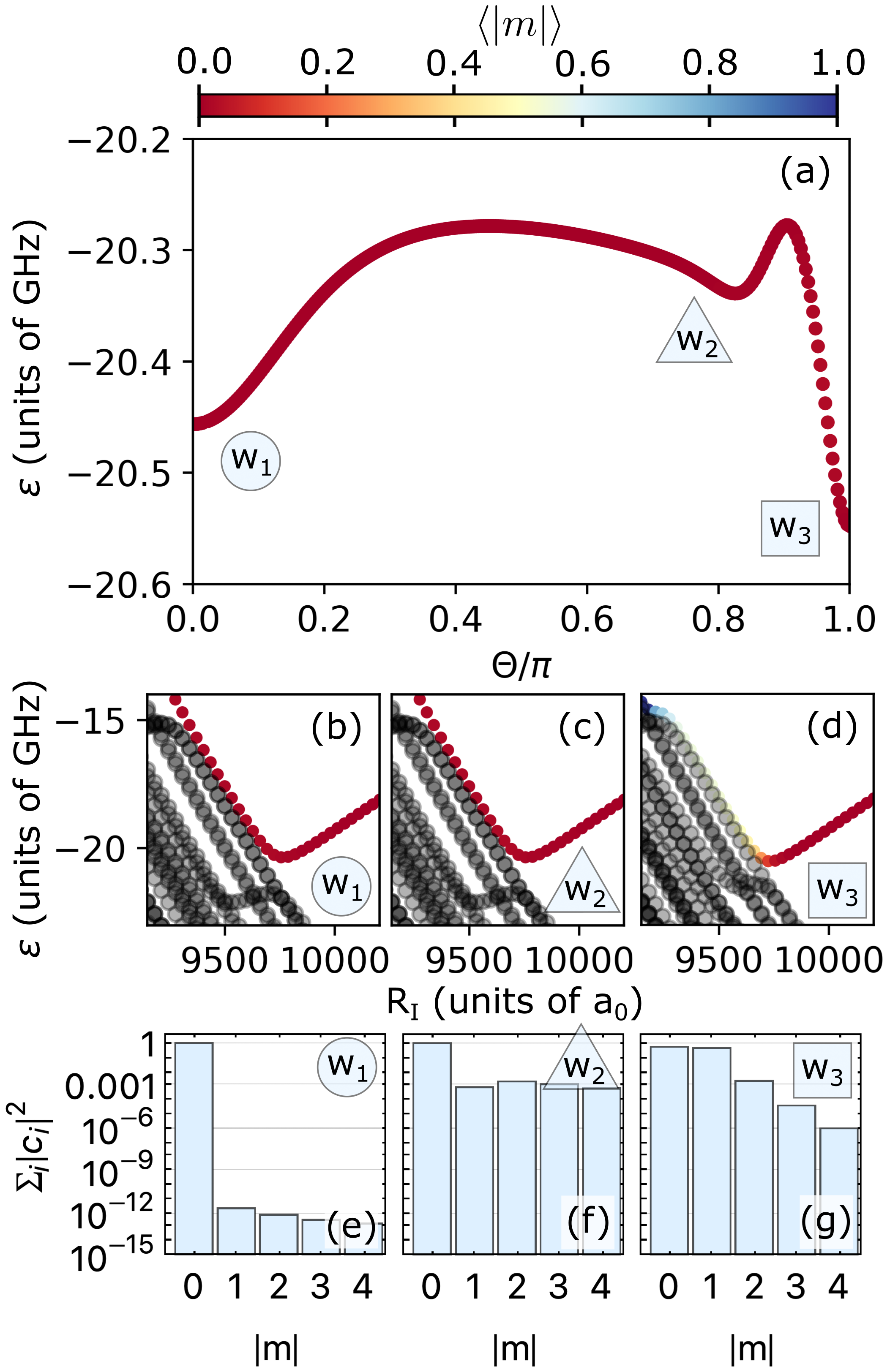}
    \caption[Squid state trimer $m$-character.]{$m$-admixture in the three-body squid state. (a) angular slice of the squid state's PES with colourbar denoting the electronic state's $m$-character $\braket{|m|}$. (b)-(d) radial slices of the PES near the w$_1$, w$_2$ and w$_3$ local minima, respectively. (e)-(g) the squid states' composition in terms of Rydberg basis states grouped by their $m$ quantum number at the w$_1$, w$_2$ and w$_3$ local minima, respectively. Energies are given relative to the field-free 32P Rydberg state.}
    \label{fig:m-char-squid}
\end{figure}
\begin{figure}[t] 
    \centering
    \includegraphics[width = 0.4\textwidth]{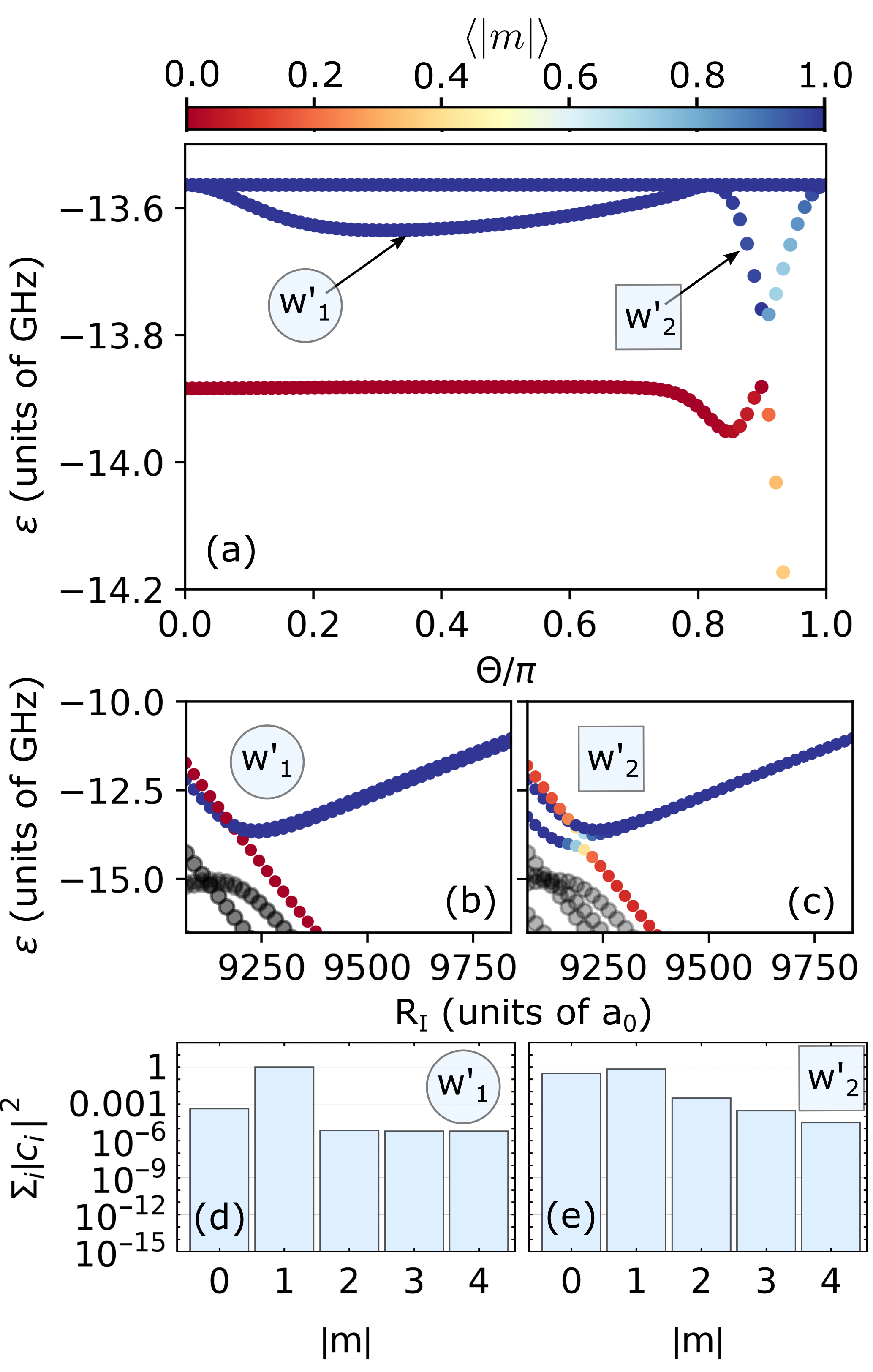}
    \caption[Snow angel state trimer $m$-character.]{$m$-admixture in the three-body snow angel state. (a) angular slice of the snow angel state's PES with colourbar denoting the electronic state's $m$-character $\braket{m}$. (b),(c) radial slices of the PES near the w$^{\prime}_1$ and w$^{\prime}_2$ local minima, respectively. (d),(f) the snow angel state's composition in terms of Rydberg basis states grouped by their $m$ quantum number at the w$_1$ and w$_2$ local minima, respectively. Energies are given relative to the field-free 32P Rydberg state.}
    \label{fig:m-char-angel}
\end{figure}
We now discuss couplings among PES of different magnetic quantum number $m$ that arise due to the presence of the ground state atom. Fig.~\ref{fig:m-char-squid} (a) shows an angular cut of the squid state's PES $\varepsilon(\Theta)$ for $R_I=9756~a_0$ and $R_A=1653~a_0$, with a colourbar denoting the $m$-character of the state. The surface exhibits potential wells at $\Theta = 0$, $\Theta = 0.8\pi$ and $\Theta = \pi$, which correspond to the local minima $\text{w}_1$, $\text{w}_2$ and $\text{w}_3$ defined in Fig.~\ref{fig:trimer-pes} (b), respectively. Along this cut, the $m$-character is pure $m=0$ and no other surfaces are present in the energy window shown. However, as is visible from Fig.~\ref{fig:dimer-pecs}, the ion-Rydberg dimer's potential well exists above a fan of high-$l$ Stark-split states. Fig.~\ref{fig:m-char-squid} (b)-(d) show slices of the PES along $R_I$ for values of $\Theta$ and $R_A$ taken around the three local minima in Fig.~\ref{fig:m-char-squid} (a). For wells $\text{w}_1$ and $\text{w}_2$, the PES are qualitatively similar to those of the dimer and show only marginal deviation from $m=0$ character. This is further illustrated by Fig.~\ref{fig:m-char-squid} (e) and (f), which show the cumulative overlap coefficients $|c_i|^2 = |\braket{\psi(R_I,R_A,\Theta)|\varphi_i}|^2$ of the squid state $\psi(R_I,R_A,\Theta)$ for each local minimum with atomic Rydberg basis states $\{\varphi_i\}$, grouped by their magnetic quantum number $m$. The lack of coupling is perhaps most surprising for the case of well w$_2$, which corresponds to a non-linear configuration of the three atoms and is hence the least symmetric of all three wells. In contrast, the slice along $R_I$ for well $\text{w}_3$ in Fig.~\ref{fig:m-char-squid} (d) shows that the local minimum is shifted to slightly lower $R_I$, such that it crosses into the fan of high-$l$ states. This observed shift does not seem to modify the $l$-character, but it does result in an admixture with states of finite $m$ at small angles around the local minimum $\Theta = \pi+\epsilon$. This admixture is primarily with $|m|=1$ states, as shown in Fig.~\ref{fig:m-char-squid} (g). Note that these couplings vanish at $\Theta=\pi$, since cylindrical symmetry is restored, i.e. the three atoms are co-linear.\\ 
As already mentioned in section~\ref{ssec:setup} in the context of the dimer system, the squid state's PEC crosses the curve of the snow angel state near the snow angel's potential well (cf. well C in Fig.~\ref{fig:dimer-pecs}). In the dimer system, this crossing is exact since there can be no coupling among surfaces of different magnetic quantum number $m$. In the trimer system, this no longer holds and a finite coupling between the $m=0$ squid surface and the $|m|=1$ snow angel surfaces is observed in Fig.~\ref{fig:m-char-angel}. This admixture is most significant near the w$^{\prime}_2$ well, for which a pronounced avoided crossing arises, as shown in Figure~\ref{fig:m-char-angel} (a) and (c). Whilst the shallower w$^{\prime}_1$ well also shows minor admixture (Figure~\ref{fig:m-char-angel} (d)), it is nevertheless orders of magnitude smaller than for w$^{\prime}_2$ and if any avoided crossing is present, its gap cannot be discerned on the energy scale of Figure~\ref{fig:m-char-angel} (b).\\
From here on, we limit our analysis of the vibrational structure to those local minima which show only relatively weak $m$-admixture (w$_1$, w$_2$ and w$^{\prime}_1$) because we expect the Born-Oppenheimer approximation to continue to hold for these. We treat crossings with states of different $m$-character near these minima as being approximately exact, though this only applies to w$^{\prime}_1$, since no crossings were observed in the vicinity of the minima of wells w$_1$ and w$_2$. For the remaining wells (w$_3$ and w$^{\prime}_2$), we expect strong non-adiabatic couplings to exist with neighbouring PES.\\

\section{Vibrational structure of the trimer}~\label{sec:timer-vib}
In this section, we explore the vibrational structure of the three-body system defined by the nuclear Hamiltonian $H_n$ in the vicinity of the local minima w$_1$, w$_2$ and w$^{\prime}_1$ of the electronic squid and snow angel states. We begin by discussing our computational approach and afterward present the analysis of the vibrational structure.\\
\subsection{Methodology and computational approach}
The total Hamiltonian of our trimer system reads $H = H_e + H_n$. In the previous section, we have discussed solutions to the electronic Hamiltonian $H_e$ obtained via exact diagonalisation in a finite basis of Rydberg states $\{\ket{n,l,m}\}$. Here, we focus on the vibrational motion of the nuclei, assuming that the system possesses zero angular momentum $J=0$. Accordingly, the full vibrational Hamiltonian~\cite{Fey2019Building,Carter1982Variational,Handy1987Derivation} is given by
\begin{widetext}
\begin{equation}\label{eq:vib_hamiltonian}
    \begin{split}
        H_n &= \frac{1}{m} \bigg[ -\frac{\partial^2}{\partial R_I^2} -\frac{\partial^2}{\partial R_A^2} -\cos(\Theta)\frac{\partial}{\partial R_I}\frac{\partial}{\partial R_A} \bigg]\\
        &-\frac{1}{m}\bigg(\frac{1}{R_I^2}+\frac{1}{R_A^2}-\frac{\cos(\Theta)}{R_I R_A} \bigg)\bigg( \frac{\partial^2}{\partial\Theta^2} +\cot(\Theta)\frac{\partial}{\partial\Theta}\bigg)\\
        &-\frac{1}{m}\bigg( \frac{1}{R_I R_A}-\frac{1}{R_A}\frac{\partial}{\partial R_I} - \frac{1}{R_I}\frac{\partial}{\partial R_A}\bigg)\bigg(\cos(\Theta) + \sin(\Theta)\frac{\partial}{\partial\Theta} \bigg)\\
        &+\varepsilon_{\nu}(R_I,R_A,\Theta),\\
    \end{split}
\end{equation}
\end{widetext}
where $m$ is the atomic mass of Rb and $\varepsilon_{\nu}(R_I,R_A,\Theta)$ is the $\nu^{\text{th}}$ adiabatic electronic PES satisfying $H_e\psi_{\nu}(\mathbf{r};R_I,R_A,\Theta)=\varepsilon_{\nu}(R_I,R_A,\Theta)\psi_{\nu}(\mathbf{r};R_I,R_A,\Theta)$. The time-independent Schr\"{o}dinger equation for the vibrational Hamiltonian in the Born-Oppenheimer approximation reads
\begin{equation}\label{eq:TISE_vib_ham}
    H_n\chi_{i}(R_I,R_A,\Theta) = E_i\chi_i(R_I,R_A,\Theta),
\end{equation}
with vibrational eigenstates $\{\chi_{i}(R_I,R_A,\Theta)\}$ and associated eigenenergies $E_i$.\\
In order to efficiently solve the above vibrational problem, we utilise the powerful Multi-Configuration Time-Dependent Hartree method (MCTDH)~\cite{mey90:73,bec00:1,mey03:251,mey06:179,dor08:224109,mey12:351,mctdh:MLpackage}. A thorough introduction to MCTDH can be found in~\cite{mey06:179,mey12:351}. In the following we provide a brief account of the approach in order to be self-contained. MCTDH is an \textit{ab initio} method for multi-mode wavepacket propagation in high-dimensional spaces. The MCTDH representation of our vibrational wavefunction $\chi(R_I,R_A,\Theta,t)$ is written as a series of Hartree products:
\begin{equation}\label{eq:mctdh1}
\begin{split}
    \chi(R_I,R_A,\Theta,t) &= \sum_{i_1=1}^{n_1}\sum_{i_2=1}^{n_2}\sum_{i_3=1}^{n_3} A_{i_1,i_2,i_3}(t)\\
    &\times \varphi_{i_1}^{(1)}(R_I,t)\varphi_{i_2}^{(2)}(R_A,t)\varphi_{i_3}^{(3)}(\Theta,t),
\end{split}
\end{equation}
where $A_{i_1,i_2,i_3}(t)$ are time-dependent coefficients and $\{\varphi_{i_d}^{(d)}\}_{i_d=1}^{n_d}$ are the so-called single particles functions (SPFs) for the $d^{\text{th}}$ degree of freedom, for which a total of $n_d$ SPFs are used. MCTDH reduces computational effort by employing a small time-dependent basis that evolves according to the Dirac-Frenkel variational principle $\braket{\delta\chi|(i\partial_t-\hat{H})|\chi}=0$. This ensures that the basis follows the active part of the complete Hilbert space as it evolves over time. The time-depdendent SPFs are described using a time-independent discrete variable representation (DVR)~\cite{Harris1965Calculation}. In this work, we use sine DVRs for the radial degrees of freedom and a Legendre DVR for the angular degree of freedom. SPFs for the radial degrees of freedom are normalised as $\int dR |\varphi(R)|^2=1$ and the angular SPFs are normalised according to $\int d\Theta \sin\Theta |\varphi(\Theta)|^2=1$.\\
The vibrational ground-state can be obtained by propagating the starting wavepacket in imaginary time, whereas excited states are obtained using so-called improved relaxation. Here, the time-dependent coefficients $\{A_{i_1,i_2,i_3}(t)\}$ are set equal to an eigenvector $\mathbf{A}$ of the Hamiltonian in the instantaneous SPF basis $\{\varphi_{i_d}^{(d)}\}_{i_d=1}^{n_d}$. The coefficients are then kept constant whilst the SPFs are relaxed using imaginary time propagation. This process is repeated until $\chi$ converges to a stationary state of the Hamiltonian. For our analysis, we seek the lowest few eigenstates in the potential wells w$_1$, w$_2$ and w$^{\prime}_1$ of the squid and snow angel PES. These are obtained by first performing a block improved relaxation scheme~\cite{mey12:351}, which provides approximate results for the first 30-40 lowest energy eigenstates. Out of this set, we select the states of interest and relax each of them individually using improved relaxation until convergence is achieved.\\
We consider a calculation to be converged when the change in energy between time steps remains consistently less than 1~Hz over a period of 100 time steps. Additionally, we assure that the occupation of the $n_d^{\text{th}}$ orbital for each degree of freedom is less than $0.1\%$ and that the occupations $\{A_{i_1,i_2,i_3}(t)\}$ decrease exponentially. For the majority of our calculations, $100$ grid points and $10$ single particle functions for each degree of freedom were sufficient to converge all eigenenergies.\\
\subsection{Vibrational structure}
\begin{figure*}[t]
    \centering
    \includegraphics[width = \figwidth]{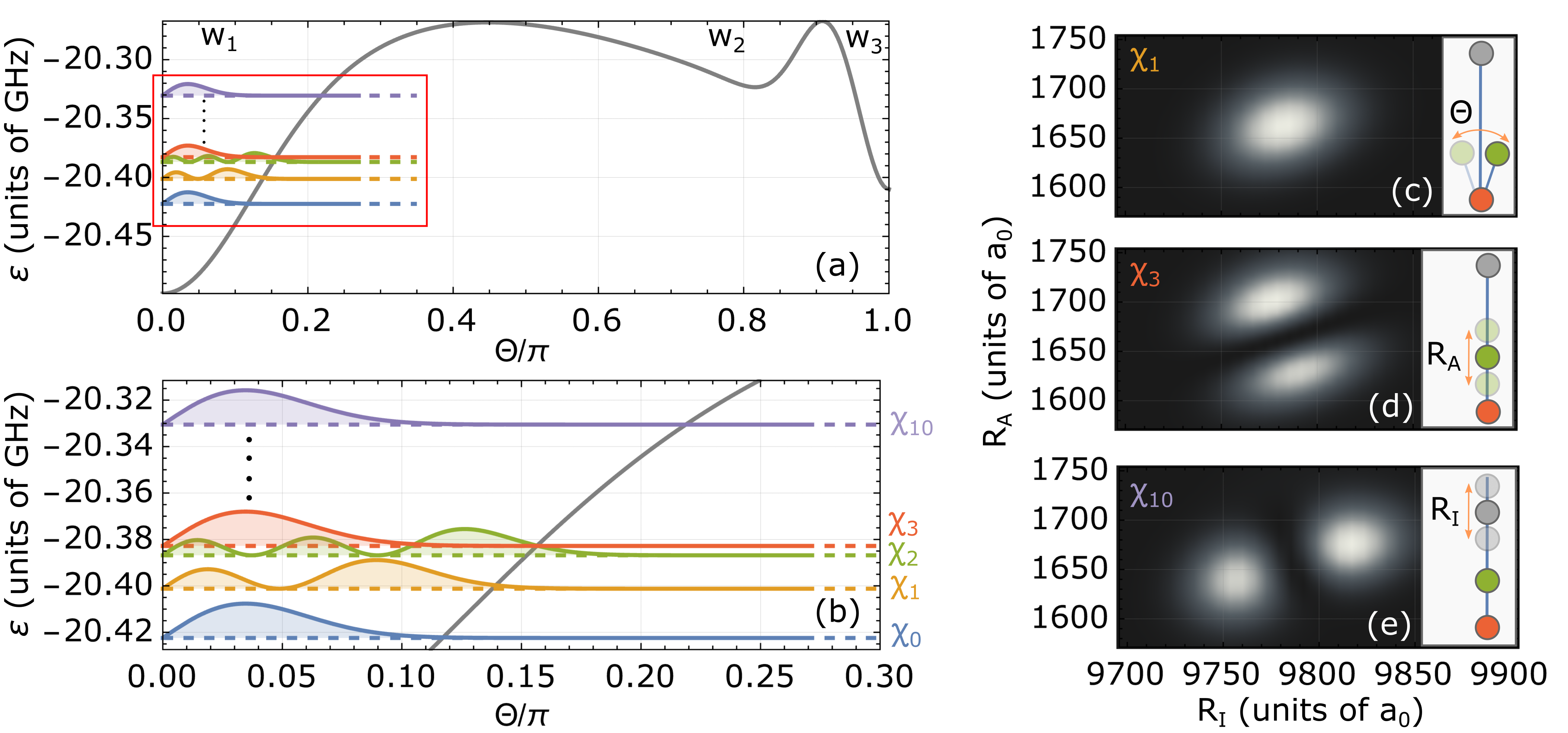}
    \caption[Lowest vibrational states of squid state trimer w$_1$ potential well.]{Selection of vibrational states of the squid state trimer's w$_1$ potential well. (a) angular cut through the potential energy surface (grey) near the local minimum $(R_I,R_A,\Theta) = (9782a_0,1657a_0,0)$. (b) reduced angular densities $\rho_{\nu}(\Theta) = \int dR_I dR_A |\chi_{\nu}(R_I,R_A,\Theta)|^2$ of the vibrational states $\nu=\{0,1,2,3,10\}$ (filled curves) offset by their energy (dashed lines). (c),(d) and (e) show reduced radial densities $\rho_{\nu}(R_I,R_A) = \int d\Theta |\chi_{\nu}(R_I,R_A,\Theta)|^2$ of $\chi_1$, $\chi_3$ and $\chi_{10}$, respectively. The eigenstates are normalised as $\int dR_I dR_A d\Theta |\chi_{\nu}(R_I,R_A,\Theta)|^2 = 1$. All energies given relative to the field-free 32P atomic Rydberg state.}
    \label{fig:vib-squid-w1}
\end{figure*}
\begin{table*}[!ht]
  \centering
  \begin{tabular}{c|c|c|c|c|c|cc}
    \toprule
    \multicolumn{1}{c|}{} & $\Delta\braket{R_I}/a_0$ & $\Delta\braket{d_z}$/Debye & $\Delta E_{000}$/MHz & $\tilde{E}_{001}$/MHz & $\tilde{E}_{010}$/MHz & $\tilde{E}_{100}$/MHz \\ \hline
    Squid $\text{w}_1$ & 14.1 & 18.2 & -181.9 & 21.2 & 39.6 & 91.9 (88.1)\\ 
    Squid $\text{w}_2$ &-5.3 & 20.8 & -102.8 & 7.4 & 31.1 & 92.6 (88.1)\\ 
    Angel $\text{w}^{\prime}_1$ & 5.9 & 19.3 & -79.9 & 3.0 & 29.3 & 86.6 (86.6)\\ \bottomrule
  \end{tabular}
  \caption[Tabular summary of vibrational state properties.]{Contrasting properties of the dimer and trimer vibrational states. From left to right, the first three columns give the difference in the vibrational ground state's (I) expected ion-Rydberg separation $\Delta\braket{R_I} = \braket{R_I^{\text{trimer}}}-\braket{R_I^{\text{dimer}}}$, (II) expected electric dipole moment of the Rydberg atom $\Delta\braket{d_z} = \braket{d_z^{\text{trimer}}} - \braket{d_z^{\text{dimer}}}$ and (III) energy $\Delta E_{000} = E_{000}^{\text{trimer}}-E_{000}^{\text{dimer}}$. The last three columns contain the trimer's excitation energies wrt. the trimer ground state of the first (IV) $\Theta$ bending mode $\tilde{E}_{001} = E_{001} - E_{000}$, (V) $R_A$ stretching mode $\tilde{E}_{010} = E_{010} - E_{000}$ and (VI) $R_I$ stretching mode $\tilde{E}_{100} = E_{100} - E_{000}$. Where applicable, corresponding values for the dimer are given in brackets. All quantities are rounded to the nearest decimal place.}
  \label{table:vib-data}
\end{table*}
\begin{figure*}[t]
    \centering
    \includegraphics[width = \figwidth]{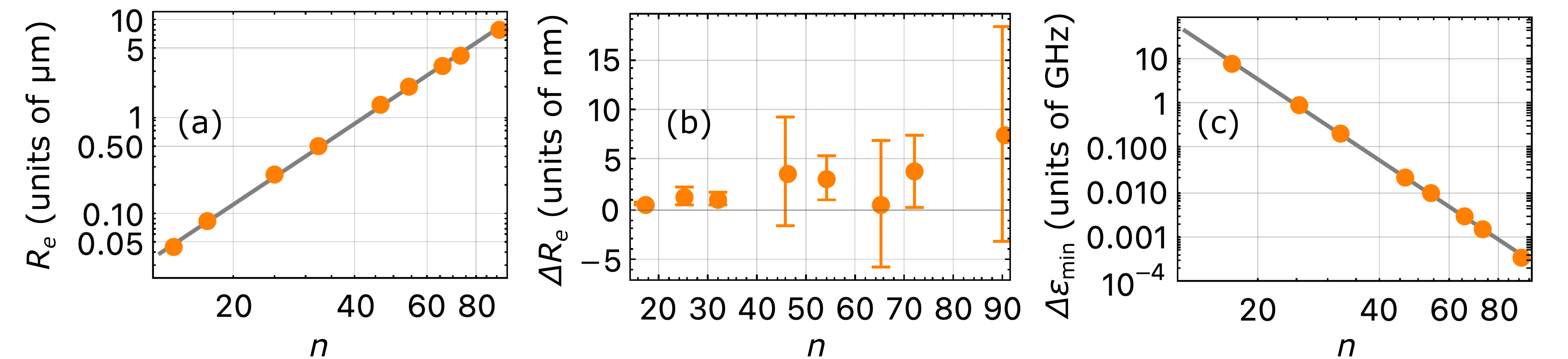}
    \caption[Comparison of squid dimer and trimer properties.]{(a) Scaling of the squid dimer's bond length with principal quantum number $n$. (b) difference in ion-Rydberg equilibrium separation and (c) potential well minima between the squid dimer and squid trimer for the w$_1$ potential well (cf. well Fig.~\ref{fig:trimer-pes}). Error bars in (b) represent the error in $\Delta R_{\text{e}}$, which arise due to the fact that the potential energy surfaces are represented on a discrete grid. Error bars for $R_{\text{e}}^{\text{dimer}}$ have been left out since their size is negligible on the scale of (a).}
    \label{fig:trimer-v-dimer}
\end{figure*}
Fig.~\ref{fig:vib-squid-w1} shows results for the reduced angular and radial densities of the four lowest vibrational states in the squid state's w$_1$ potential well. Table~\ref{table:vib-data} summarises properties of the lowest vibrational states in all three trimer wells and compares these with the vibrational states of the squid and snow angel dimer.\\
In Fig.~\ref{fig:vib-squid-w1} (a) and (b), we observe that the energetically lowest states in the potential well w$_1$ are localised around $\Theta=0$ such that the three atoms are arranged along a common internuclear axis, i.e. in a collinear configuration. The well's ground state is approximately 180~MHz lower in energy than the ground state of the squid state dimer and exhibits a positive shift of the ion-Rydberg bond length and corresponding electric dipole moment, as shown in Table~\ref{table:vib-data}. From Fig.~\ref{fig:vib-squid-w1} (b), we find that the reduced angular density of the first excited state exhibits an additional node, indicating that this is a bending mode. The excitation energy of this mode is about 20~MHz above the ground state energy. In contrast, the first excited state of the dimer is approximately 88.1~MHz. The second excited state is also a bending mode, with an additional node along the reduced angular density in Fig.~\ref{fig:vib-squid-w1} (b). The third excited state's energy lies close to that of the second excited state, yet its reduced radial density in Fig.~\ref{fig:vib-squid-w1} (d) confirms that it is an excitation in $R_A$, a stretching mode. The energy gap to the first excitation in $R_I$ is 91.6~MHz, which is of the same order of magnitude as the energy gap between vibrational states of the dimer.\\
We now examine some global trends of properties among the three wells evident from Table~\ref{table:vib-data}. Surprisingly, the ground state atom's presence appears to contribute to a consistent positive shift of 20 Debye in the Rydberg atom's electric dipole moment $\braket{d_z}$, independent of which well it inhabits and regardless of whether the expected ion-Rydberg binding length $\braket{R_I}$ has increased or decreased with respect to the dimer. On the other hand, the energy gap to the first $R_I$ stretching mode is essentially unchanged by the presence of the perturber.
The energy gaps of the bending and $R_A$ stretching modes do vary among the different potential wells. In particular, in the w$^{\prime}_1$ well 3~MHz are necessary to excite its first bending mode, whilst the same gap in the w$_1$ well is an order of magnitude larger. From the 2D graphs of the PES in Fig.~\ref{fig:trimer-pes} (b) and (d), we see that w$_1$ and w$^{\prime}_1$ are reasonably extended along $\Theta$ which implies a weak angular confinement in both cases. Nevertheless, the w$_1$ well is considerably deeper than the w$^{\prime}_1$ well which leads the greater excitation gap.\\ 
In an experimental setting, it should be possible to distinguish trimer states from dimer states due to the trimer's deeper binding energies and distinguishable excitation series for bending and stretching modes depending on the geometric configuration of the nuclei. Moreover, mass spectroscopy techniques such as that employed in~\cite{Zuber2022Observation} could be used to distinguish the two based on their mass difference. In principle, the trimer's shift in ion-Rydberg binding length could serve as a further indicator for the presence of trimer states. The shifts predicted in Table~\ref{table:vib-data} are however on the order of a few Bohr radii, which is clearly too small to be observed with current techniques. As first shown in~\cite{Deiss2021LongRange}, the equilibrium separation of the ion-Rydberg dimer $R_{\text{e}}$ scales with the principal quantum number $n$ and in Fig.~\ref{fig:trimer-v-dimer} (a) we find this scaling to be $R_{\text{e}} \propto n^{2.80\pm0.18}$, which is close to the value expected from perturbation theory as described in~\ref{ssec:dimer}. 
However, we see no clear trend for the change in the equilibrium separation $\Delta R_{\text{e}} = R_{\text{e}}^{\text{trimer}}-R_{\text{e}}^{\text{dimer}}$ for the well w$_1$, shown in Fig.~\ref{fig:trimer-v-dimer} (b).\\
On the other hand, there is an evident trend in the difference of the local energy minima $\Delta \epsilon_{\text{min}} = \epsilon_{\text{min}}^{\text{dimer}} - \epsilon_{\text{min}}^{\text{trimer}}$ between the PES of the dimer and trimer. Fig.~\ref{fig:trimer-v-dimer} (c) shows $\Delta \epsilon_{\text{min}}$ for the trimer w$_1$ well and we find $\Delta \epsilon_{\text{min}}\propto n^{-5.97\pm0.06}$. This can be understood as follows. To first order, the interaction of the ground state atom with the Rydberg electron contributes an energy shift proportional to the electron's probability density $\Delta \epsilon \propto |\psi(R_A)|^2$, whose sign depends on the s-wave scattering length $a_s[k(R_A)]$. The electron's probability density scales with the radius of the Rydberg orbit as $|\psi(r)|^2 \propto 1/r^3$, which in turn scales quadratically with $n$. This gives $\Delta \epsilon \propto n^{-6}$, which is close to our fitted value. Since $\Delta \epsilon_{\text{min}}$ decays with $n$, the trimer's vibrational spectrum will lie energetically closer to the dimer's at higher principal quantum numbers, making them harder to distinguish. By modulating the density of the background gas of atoms, signals of trimer states can be suppressed or enhanced.

\section{Conclusions}\label{sec:summary}
In this work, we first studied the case of an ion interacting with a Rydberg atom and determined the electronic structure of bound states which exist in various potential wells in the spectrum of adiabatic potential energy curves. For atomic species like Rb and Cs, the ion's presence mixes the quantum defect-split $p$-state with its neighbouring quasi-degenerate manifold of high-$l$ states. This results in the formation of a twofold series of potential wells, one for each $|m|$ sub-level of the $p$-state. We found that the Rydberg electron's probability density exhibits several lobes, which vary in their size and number between the wells. Due to their distinctive shapes, we classify the electronic density patterns along the $m=0$ series as squid states, whilst those along the $|m|=1$ series are termed snow angel states.\\
We predict that the lobes in the electronic density admit the binding of additional ground state atoms via the same binding mechanism as conventional polar and non-polar ULRM. We have demonstrated this for a three-body system of a Rydberg atom interacting with a ground state atom and a single ion, in the limit of non-overlapping charge distributions. Bound vibrational states exist among the three species, forming a charged ultralong-range Rydberg trimer. The ground state atom can become bound within different lobes in the electronic density, such that linear as well as non-linear geometric configurations of the nuclear framework are possible.\\
The ground state atom leads to a significant shift in the energy of the trimer's vibrational ground state and the additional nuclear degrees of freedom give rise to distinct excitation series for bending and stretching modes, which vary between the local minima associated with different lobes in the electronic structure. Using current spectroscopic techniques, it should thus be possible to distinguish the trimer and dimer states. Similarly to the ion-Rydberg dimer, we expect nonadiabatic couplings between neighbouring potential energy surfaces to be a significant channel for molecular decay in our system. Indeed, the effect may even be stronger in the trimer system since crossings between states of different $m$ are no longer symmetry-protected. Corresponding investigations are left to future studies.\\
Our work lays the foundation for exploring ULRM in inhomogeneous electric fields. Future work could consider the interaction of an ion with polyatomic ULRM and wavepacket dynamics in the ion's field. Additionally, conical intersections can occur between adiabatic potential energy surfaces in triatomic systems and thus one could examine the role of beyond Born-Oppenheimer physics in our system in the spirit of recent work in this direction~\cite{Hummel2021Synthetic}.\\

\section*{Acknowledgements}
D. J. B. thanks Maxim Pyzh for many fruitful discussions. This work is funded by the Cluster of Excellence “Advanced Imaging of Matter” of the Deutsche Forschungsgemeinschaft (DFG)-EXC 2056, Project ID No. 390715994.

\bibliography{ChargedTrimer,mctdh}

\begin{thebibliography}{76}%
\makeatletter
\providecommand \@ifxundefined [1]{%
 \@ifx{#1\undefined}
}%
\providecommand \@ifnum [1]{%
 \ifnum #1\expandafter \@firstoftwo
 \else \expandafter \@secondoftwo
 \fi
}%
\providecommand \@ifx [1]{%
 \ifx #1\expandafter \@firstoftwo
 \else \expandafter \@secondoftwo
 \fi
}%
\providecommand \natexlab [1]{#1}%
\providecommand \enquote  [1]{``#1''}%
\providecommand \bibnamefont  [1]{#1}%
\providecommand \bibfnamefont [1]{#1}%
\providecommand \citenamefont [1]{#1}%
\providecommand \href@noop [0]{\@secondoftwo}%
\providecommand \href [0]{\begingroup \@sanitize@url \@href}%
\providecommand \@href[1]{\@@startlink{#1}\@@href}%
\providecommand \@@href[1]{\endgroup#1\@@endlink}%
\providecommand \@sanitize@url [0]{\catcode `\\12\catcode `\$12\catcode
  `\&12\catcode `\#12\catcode `\^12\catcode `\_12\catcode `\%12\relax}%
\providecommand \@@startlink[1]{}%
\providecommand \@@endlink[0]{}%
\providecommand \url  [0]{\begingroup\@sanitize@url \@url }%
\providecommand \@url [1]{\endgroup\@href {#1}{\urlprefix }}%
\providecommand \urlprefix  [0]{URL }%
\providecommand \Eprint [0]{\href }%
\providecommand \doibase [0]{http://dx.doi.org/}%
\providecommand \selectlanguage [0]{\@gobble}%
\providecommand \bibinfo  [0]{\@secondoftwo}%
\providecommand \bibfield  [0]{\@secondoftwo}%
\providecommand \translation [1]{[#1]}%
\providecommand \BibitemOpen [0]{}%
\providecommand \bibitemStop [0]{}%
\providecommand \bibitemNoStop [0]{.\EOS\space}%
\providecommand \EOS [0]{\spacefactor3000\relax}%
\providecommand \BibitemShut  [1]{\csname bibitem#1\endcsname}%
\let\auto@bib@innerbib\@empty
\bibitem [{\citenamefont {Zipkes}\ \emph {et~al.}(2011)\citenamefont {Zipkes},
  \citenamefont {Ratschbacher}, \citenamefont {Palzer}, \citenamefont {Sias},\
  and\ \citenamefont {K{\"o}hl}}]{Zipkes2011Hybrid}%
  \BibitemOpen
  \bibfield  {author} {\bibinfo {author} {\bibfnamefont {C.}~\bibnamefont
  {Zipkes}}, \bibinfo {author} {\bibfnamefont {L.}~\bibnamefont
  {Ratschbacher}}, \bibinfo {author} {\bibfnamefont {S.}~\bibnamefont
  {Palzer}}, \bibinfo {author} {\bibfnamefont {C.}~\bibnamefont {Sias}}, \ and\
  \bibinfo {author} {\bibfnamefont {M.}~\bibnamefont {K{\"o}hl}},\ }\href
  {\doibase 10.1088/1742-6596/264/1/012019} {\bibfield  {journal} {\bibinfo
  {journal} {J. Phys.: Conf. Ser.}\ }\textbf {\bibinfo {volume} {264}},\
  \bibinfo {pages} {012019} (\bibinfo {year} {2011})}\BibitemShut {NoStop}%
\bibitem [{\citenamefont {Tomza}\ \emph {et~al.}(2019)\citenamefont {Tomza},
  \citenamefont {Jachymski}, \citenamefont {Gerritsma}, \citenamefont
  {Negretti}, \citenamefont {Calarco}, \citenamefont {Idziaszek},\ and\
  \citenamefont {Julienne}}]{Tomza2019Cold}%
  \BibitemOpen
  \bibfield  {author} {\bibinfo {author} {\bibfnamefont {M.}~\bibnamefont
  {Tomza}}, \bibinfo {author} {\bibfnamefont {K.}~\bibnamefont {Jachymski}},
  \bibinfo {author} {\bibfnamefont {R.}~\bibnamefont {Gerritsma}}, \bibinfo
  {author} {\bibfnamefont {A.}~\bibnamefont {Negretti}}, \bibinfo {author}
  {\bibfnamefont {T.}~\bibnamefont {Calarco}}, \bibinfo {author} {\bibfnamefont
  {Z.}~\bibnamefont {Idziaszek}}, \ and\ \bibinfo {author} {\bibfnamefont
  {P.~S.}\ \bibnamefont {Julienne}},\ }\href {\doibase
  10.1103/RevModPhys.91.035001} {\bibfield  {journal} {\bibinfo  {journal}
  {Rev. Mod. Phys.}\ }\textbf {\bibinfo {volume} {91}},\ \bibinfo {pages}
  {035001} (\bibinfo {year} {2019})}\BibitemShut {NoStop}%
\bibitem [{\citenamefont {Lous}\ and\ \citenamefont
  {Gerritsma}(2022)}]{Lous2022Chapter}%
  \BibitemOpen
  \bibfield  {author} {\bibinfo {author} {\bibfnamefont {R.~S.}\ \bibnamefont
  {Lous}}\ and\ \bibinfo {author} {\bibfnamefont {R.}~\bibnamefont
  {Gerritsma}},\ }in\ \href {\doibase 10.1016/bs.aamop.2022.05.002} {\emph
  {\bibinfo {booktitle} {Advances {{In Atomic}}, {{Molecular}}, and {{Optical
  Physics}}}}},\ \bibinfo {series} {Advances in {{Atomic}}, {{Molecular}}, and
  {{Optical Physics}}}, Vol.~\bibinfo {volume} {71},\ \bibinfo {editor} {edited
  by\ \bibinfo {editor} {\bibfnamefont {L.~F.}\ \bibnamefont {DiMauro}},
  \bibinfo {editor} {\bibfnamefont {H.}~\bibnamefont {Perrin}}, \ and\ \bibinfo
  {editor} {\bibfnamefont {S.~F.}\ \bibnamefont {Yelin}}}\ (\bibinfo
  {publisher} {{Academic Press}},\ \bibinfo {year} {2022})\ pp.\ \bibinfo
  {pages} {65--133}\BibitemShut {NoStop}%
\bibitem [{\citenamefont {Ratschbacher}\ \emph {et~al.}(2012)\citenamefont
  {Ratschbacher}, \citenamefont {Zipkes}, \citenamefont {Sias},\ and\
  \citenamefont {K{\"o}hl}}]{Ratschbacher2012Controlling}%
  \BibitemOpen
  \bibfield  {author} {\bibinfo {author} {\bibfnamefont {L.}~\bibnamefont
  {Ratschbacher}}, \bibinfo {author} {\bibfnamefont {C.}~\bibnamefont
  {Zipkes}}, \bibinfo {author} {\bibfnamefont {C.}~\bibnamefont {Sias}}, \ and\
  \bibinfo {author} {\bibfnamefont {M.}~\bibnamefont {K{\"o}hl}},\ }\href
  {\doibase 10.1038/nphys2373} {\bibfield  {journal} {\bibinfo  {journal}
  {Nature Phys}\ }\textbf {\bibinfo {volume} {8}},\ \bibinfo {pages} {649}
  (\bibinfo {year} {2012})}\BibitemShut {NoStop}%
\bibitem [{\citenamefont {Hall}\ and\ \citenamefont
  {Willitsch}(2012)}]{Hall2012Millikelvin}%
  \BibitemOpen
  \bibfield  {author} {\bibinfo {author} {\bibfnamefont {F.~H.~J.}\
  \bibnamefont {Hall}}\ and\ \bibinfo {author} {\bibfnamefont {S.}~\bibnamefont
  {Willitsch}},\ }\href {\doibase 10.1103/PhysRevLett.109.233202} {\bibfield
  {journal} {\bibinfo  {journal} {Phys. Rev. Lett.}\ }\textbf {\bibinfo
  {volume} {109}},\ \bibinfo {pages} {233202} (\bibinfo {year}
  {2012})}\BibitemShut {NoStop}%
\bibitem [{\citenamefont {H{\"a}rter}\ \emph {et~al.}(2012)\citenamefont
  {H{\"a}rter}, \citenamefont {Kr{\"u}kow}, \citenamefont {Brunner},
  \citenamefont {Schnitzler}, \citenamefont {Schmid},\ and\ \citenamefont
  {Denschlag}}]{Harter2012Single}%
  \BibitemOpen
  \bibfield  {author} {\bibinfo {author} {\bibfnamefont {A.}~\bibnamefont
  {H{\"a}rter}}, \bibinfo {author} {\bibfnamefont {A.}~\bibnamefont
  {Kr{\"u}kow}}, \bibinfo {author} {\bibfnamefont {A.}~\bibnamefont {Brunner}},
  \bibinfo {author} {\bibfnamefont {W.}~\bibnamefont {Schnitzler}}, \bibinfo
  {author} {\bibfnamefont {S.}~\bibnamefont {Schmid}}, \ and\ \bibinfo {author}
  {\bibfnamefont {J.~H.}\ \bibnamefont {Denschlag}},\ }\href {\doibase
  10.1103/PhysRevLett.109.123201} {\bibfield  {journal} {\bibinfo  {journal}
  {Phys. Rev. Lett.}\ }\textbf {\bibinfo {volume} {109}},\ \bibinfo {pages}
  {123201} (\bibinfo {year} {2012})}\BibitemShut {NoStop}%
\bibitem [{\citenamefont {Meir}\ \emph {et~al.}(2016)\citenamefont {Meir},
  \citenamefont {Sikorsky}, \citenamefont {{Ben-shlomi}}, \citenamefont
  {Akerman}, \citenamefont {Dallal},\ and\ \citenamefont
  {Ozeri}}]{Meir2016Dynamics}%
  \BibitemOpen
  \bibfield  {author} {\bibinfo {author} {\bibfnamefont {Z.}~\bibnamefont
  {Meir}}, \bibinfo {author} {\bibfnamefont {T.}~\bibnamefont {Sikorsky}},
  \bibinfo {author} {\bibfnamefont {R.}~\bibnamefont {{Ben-shlomi}}}, \bibinfo
  {author} {\bibfnamefont {N.}~\bibnamefont {Akerman}}, \bibinfo {author}
  {\bibfnamefont {Y.}~\bibnamefont {Dallal}}, \ and\ \bibinfo {author}
  {\bibfnamefont {R.}~\bibnamefont {Ozeri}},\ }\href {\doibase
  10.1103/PhysRevLett.117.243401} {\bibfield  {journal} {\bibinfo  {journal}
  {Phys. Rev. Lett.}\ }\textbf {\bibinfo {volume} {117}},\ \bibinfo {pages}
  {243401} (\bibinfo {year} {2016})}\BibitemShut {NoStop}%
\bibitem [{\citenamefont {{P{\'e}rez-R{\'i}os}}(2021)}]{Perez-Rios2021Cold}%
  \BibitemOpen
  \bibfield  {author} {\bibinfo {author} {\bibfnamefont {J.}~\bibnamefont
  {{P{\'e}rez-R{\'i}os}}},\ }\href {\doibase 10.1080/00268976.2021.1881637}
  {\bibfield  {journal} {\bibinfo  {journal} {Mol. Phys.}\ }\textbf {\bibinfo
  {volume} {119}},\ \bibinfo {pages} {e1881637} (\bibinfo {year}
  {2021})}\BibitemShut {NoStop}%
\bibitem [{\citenamefont {Oghittu}\ and\ \citenamefont
  {Negretti}(2022)}]{Oghittu2022Quantumlimited}%
  \BibitemOpen
  \bibfield  {author} {\bibinfo {author} {\bibfnamefont {L.}~\bibnamefont
  {Oghittu}}\ and\ \bibinfo {author} {\bibfnamefont {A.}~\bibnamefont
  {Negretti}},\ }\href {\doibase 10.1103/PhysRevResearch.4.023069} {\bibfield
  {journal} {\bibinfo  {journal} {Phys. Rev. Research}\ }\textbf {\bibinfo
  {volume} {4}},\ \bibinfo {pages} {023069} (\bibinfo {year}
  {2022})}\BibitemShut {NoStop}%
\bibitem [{\citenamefont {C{\^o}t{\'e}}\ \emph {et~al.}(2002)\citenamefont
  {C{\^o}t{\'e}}, \citenamefont {Kharchenko},\ and\ \citenamefont
  {Lukin}}]{Cote2002Mesoscopic}%
  \BibitemOpen
  \bibfield  {author} {\bibinfo {author} {\bibfnamefont {R.}~\bibnamefont
  {C{\^o}t{\'e}}}, \bibinfo {author} {\bibfnamefont {V.}~\bibnamefont
  {Kharchenko}}, \ and\ \bibinfo {author} {\bibfnamefont {M.~D.}\ \bibnamefont
  {Lukin}},\ }\href {\doibase 10.1103/PhysRevLett.89.093001} {\bibfield
  {journal} {\bibinfo  {journal} {Phys. Rev. Lett.}\ }\textbf {\bibinfo
  {volume} {89}},\ \bibinfo {pages} {093001} (\bibinfo {year}
  {2002})}\BibitemShut {NoStop}%
\bibitem [{\citenamefont {Schurer}\ \emph {et~al.}(2017)\citenamefont
  {Schurer}, \citenamefont {Negretti},\ and\ \citenamefont
  {Schmelcher}}]{Schurer2017Unraveling}%
  \BibitemOpen
  \bibfield  {author} {\bibinfo {author} {\bibfnamefont {J.~M.}\ \bibnamefont
  {Schurer}}, \bibinfo {author} {\bibfnamefont {A.}~\bibnamefont {Negretti}}, \
  and\ \bibinfo {author} {\bibfnamefont {P.}~\bibnamefont {Schmelcher}},\
  }\href {\doibase 10.1103/PhysRevLett.119.063001} {\bibfield  {journal}
  {\bibinfo  {journal} {Phys. Rev. Lett.}\ }\textbf {\bibinfo {volume} {119}},\
  \bibinfo {pages} {063001} (\bibinfo {year} {2017})}\BibitemShut {NoStop}%
\bibitem [{\citenamefont {Bosworth}\ \emph {et~al.}(2021)\citenamefont
  {Bosworth}, \citenamefont {Pyzh},\ and\ \citenamefont
  {Schmelcher}}]{Bosworth2021Spectral}%
  \BibitemOpen
  \bibfield  {author} {\bibinfo {author} {\bibfnamefont {D.~J.}\ \bibnamefont
  {Bosworth}}, \bibinfo {author} {\bibfnamefont {M.}~\bibnamefont {Pyzh}}, \
  and\ \bibinfo {author} {\bibfnamefont {P.}~\bibnamefont {Schmelcher}},\
  }\href {\doibase 10.1103/PhysRevA.103.033303} {\bibfield  {journal} {\bibinfo
   {journal} {Phys. Rev. A}\ }\textbf {\bibinfo {volume} {103}},\ \bibinfo
  {pages} {033303} (\bibinfo {year} {2021})}\BibitemShut {NoStop}%
\bibitem [{\citenamefont {Schmid}\ \emph {et~al.}(2010)\citenamefont {Schmid},
  \citenamefont {H{\"a}rter},\ and\ \citenamefont
  {Denschlag}}]{Schmid2010Dynamics}%
  \BibitemOpen
  \bibfield  {author} {\bibinfo {author} {\bibfnamefont {S.}~\bibnamefont
  {Schmid}}, \bibinfo {author} {\bibfnamefont {A.}~\bibnamefont {H{\"a}rter}},
  \ and\ \bibinfo {author} {\bibfnamefont {J.~H.}\ \bibnamefont {Denschlag}},\
  }\href {\doibase 10.1103/PhysRevLett.105.133202} {\bibfield  {journal}
  {\bibinfo  {journal} {Phys. Rev. Lett.}\ }\textbf {\bibinfo {volume} {105}},\
  \bibinfo {pages} {133202} (\bibinfo {year} {2010})}\BibitemShut {NoStop}%
\bibitem [{\citenamefont {Zipkes}\ \emph {et~al.}(2010)\citenamefont {Zipkes},
  \citenamefont {Palzer}, \citenamefont {Sias},\ and\ \citenamefont
  {K{\"o}hl}}]{Zipkes2010Trapped}%
  \BibitemOpen
  \bibfield  {author} {\bibinfo {author} {\bibfnamefont {C.}~\bibnamefont
  {Zipkes}}, \bibinfo {author} {\bibfnamefont {S.}~\bibnamefont {Palzer}},
  \bibinfo {author} {\bibfnamefont {C.}~\bibnamefont {Sias}}, \ and\ \bibinfo
  {author} {\bibfnamefont {M.}~\bibnamefont {K{\"o}hl}},\ }\href {\doibase
  10.1038/nature08865} {\bibfield  {journal} {\bibinfo  {journal} {Nature}\
  }\textbf {\bibinfo {volume} {464}},\ \bibinfo {pages} {388} (\bibinfo {year}
  {2010})}\BibitemShut {NoStop}%
\bibitem [{\citenamefont {Veit}\ \emph {et~al.}(2021)\citenamefont {Veit},
  \citenamefont {Zuber}, \citenamefont {{Herrera-Sancho}}, \citenamefont
  {Anasuri}, \citenamefont {Schmid}, \citenamefont {Meinert}, \citenamefont
  {L{\"o}w},\ and\ \citenamefont {Pfau}}]{Veit2021Pulsed}%
  \BibitemOpen
  \bibfield  {author} {\bibinfo {author} {\bibfnamefont {C.}~\bibnamefont
  {Veit}}, \bibinfo {author} {\bibfnamefont {N.}~\bibnamefont {Zuber}},
  \bibinfo {author} {\bibfnamefont {O.~A.}\ \bibnamefont {{Herrera-Sancho}}},
  \bibinfo {author} {\bibfnamefont {V.~S.~V.}\ \bibnamefont {Anasuri}},
  \bibinfo {author} {\bibfnamefont {T.}~\bibnamefont {Schmid}}, \bibinfo
  {author} {\bibfnamefont {F.}~\bibnamefont {Meinert}}, \bibinfo {author}
  {\bibfnamefont {R.}~\bibnamefont {L{\"o}w}}, \ and\ \bibinfo {author}
  {\bibfnamefont {T.}~\bibnamefont {Pfau}},\ }\href {\doibase
  10.1103/PhysRevX.11.011036} {\bibfield  {journal} {\bibinfo  {journal} {Phys.
  Rev. X}\ }\textbf {\bibinfo {volume} {11}},\ \bibinfo {pages} {011036}
  (\bibinfo {year} {2021})}\BibitemShut {NoStop}%
\bibitem [{\citenamefont {Gerritsma}\ \emph {et~al.}(2012)\citenamefont
  {Gerritsma}, \citenamefont {Negretti}, \citenamefont {Doerk}, \citenamefont
  {Idziaszek}, \citenamefont {Calarco},\ and\ \citenamefont
  {{Schmidt-Kaler}}}]{Gerritsma2012Bosonic}%
  \BibitemOpen
  \bibfield  {author} {\bibinfo {author} {\bibfnamefont {R.}~\bibnamefont
  {Gerritsma}}, \bibinfo {author} {\bibfnamefont {A.}~\bibnamefont {Negretti}},
  \bibinfo {author} {\bibfnamefont {H.}~\bibnamefont {Doerk}}, \bibinfo
  {author} {\bibfnamefont {Z.}~\bibnamefont {Idziaszek}}, \bibinfo {author}
  {\bibfnamefont {T.}~\bibnamefont {Calarco}}, \ and\ \bibinfo {author}
  {\bibfnamefont {F.}~\bibnamefont {{Schmidt-Kaler}}},\ }\href {\doibase
  10.1103/PhysRevLett.109.080402} {\bibfield  {journal} {\bibinfo  {journal}
  {Phys. Rev. Lett.}\ }\textbf {\bibinfo {volume} {109}},\ \bibinfo {pages}
  {080402} (\bibinfo {year} {2012})}\BibitemShut {NoStop}%
\bibitem [{\citenamefont {Schurer}\ \emph {et~al.}(2016)\citenamefont
  {Schurer}, \citenamefont {Gerritsma}, \citenamefont {Schmelcher},\ and\
  \citenamefont {Negretti}}]{Schurer2016Impact}%
  \BibitemOpen
  \bibfield  {author} {\bibinfo {author} {\bibfnamefont {J.~M.}\ \bibnamefont
  {Schurer}}, \bibinfo {author} {\bibfnamefont {R.}~\bibnamefont {Gerritsma}},
  \bibinfo {author} {\bibfnamefont {P.}~\bibnamefont {Schmelcher}}, \ and\
  \bibinfo {author} {\bibfnamefont {A.}~\bibnamefont {Negretti}},\ }\href
  {\doibase 10.1103/PhysRevA.93.063602} {\bibfield  {journal} {\bibinfo
  {journal} {Phys. Rev. A}\ }\textbf {\bibinfo {volume} {93}},\ \bibinfo
  {pages} {063602} (\bibinfo {year} {2016})}\BibitemShut {NoStop}%
\bibitem [{\citenamefont {Feldker}\ \emph {et~al.}(2020)\citenamefont
  {Feldker}, \citenamefont {F{\"u}rst}, \citenamefont {Hirzler}, \citenamefont
  {Ewald}, \citenamefont {Mazzanti}, \citenamefont {Wiater}, \citenamefont
  {Tomza},\ and\ \citenamefont {Gerritsma}}]{Feldker2020Buffer}%
  \BibitemOpen
  \bibfield  {author} {\bibinfo {author} {\bibfnamefont {T.}~\bibnamefont
  {Feldker}}, \bibinfo {author} {\bibfnamefont {H.}~\bibnamefont {F{\"u}rst}},
  \bibinfo {author} {\bibfnamefont {H.}~\bibnamefont {Hirzler}}, \bibinfo
  {author} {\bibfnamefont {N.~V.}\ \bibnamefont {Ewald}}, \bibinfo {author}
  {\bibfnamefont {M.}~\bibnamefont {Mazzanti}}, \bibinfo {author}
  {\bibfnamefont {D.}~\bibnamefont {Wiater}}, \bibinfo {author} {\bibfnamefont
  {M.}~\bibnamefont {Tomza}}, \ and\ \bibinfo {author} {\bibfnamefont
  {R.}~\bibnamefont {Gerritsma}},\ }\href {\doibase 10.1038/s41567-019-0772-5}
  {\bibfield  {journal} {\bibinfo  {journal} {Nat. Phys.}\ }\textbf {\bibinfo
  {volume} {16}},\ \bibinfo {pages} {413} (\bibinfo {year} {2020})}\BibitemShut
  {NoStop}%
\bibitem [{\citenamefont {Weckesser}\ \emph {et~al.}(2021)\citenamefont
  {Weckesser}, \citenamefont {Thielemann}, \citenamefont {Wiater},
  \citenamefont {Wojciechowska}, \citenamefont {Karpa}, \citenamefont
  {Jachymski}, \citenamefont {Tomza}, \citenamefont {Walker},\ and\
  \citenamefont {Schaetz}}]{Weckesser2021Observation}%
  \BibitemOpen
  \bibfield  {author} {\bibinfo {author} {\bibfnamefont {P.}~\bibnamefont
  {Weckesser}}, \bibinfo {author} {\bibfnamefont {F.}~\bibnamefont
  {Thielemann}}, \bibinfo {author} {\bibfnamefont {D.}~\bibnamefont {Wiater}},
  \bibinfo {author} {\bibfnamefont {A.}~\bibnamefont {Wojciechowska}}, \bibinfo
  {author} {\bibfnamefont {L.}~\bibnamefont {Karpa}}, \bibinfo {author}
  {\bibfnamefont {K.}~\bibnamefont {Jachymski}}, \bibinfo {author}
  {\bibfnamefont {M.}~\bibnamefont {Tomza}}, \bibinfo {author} {\bibfnamefont
  {T.}~\bibnamefont {Walker}}, \ and\ \bibinfo {author} {\bibfnamefont
  {T.}~\bibnamefont {Schaetz}},\ }\href {\doibase 10.1038/s41586-021-04112-y}
  {\bibfield  {journal} {\bibinfo  {journal} {Nature}\ }\textbf {\bibinfo
  {volume} {600}},\ \bibinfo {pages} {429} (\bibinfo {year}
  {2021})}\BibitemShut {NoStop}%
\bibitem [{\citenamefont {Secker}\ \emph {et~al.}(2016)\citenamefont {Secker},
  \citenamefont {Gerritsma}, \citenamefont {Glaetzle},\ and\ \citenamefont
  {Negretti}}]{Secker2016Controlled}%
  \BibitemOpen
  \bibfield  {author} {\bibinfo {author} {\bibfnamefont {T.}~\bibnamefont
  {Secker}}, \bibinfo {author} {\bibfnamefont {R.}~\bibnamefont {Gerritsma}},
  \bibinfo {author} {\bibfnamefont {A.~W.}\ \bibnamefont {Glaetzle}}, \ and\
  \bibinfo {author} {\bibfnamefont {A.}~\bibnamefont {Negretti}},\ }\href
  {\doibase 10.1103/PhysRevA.94.013420} {\bibfield  {journal} {\bibinfo
  {journal} {Phys. Rev. A}\ }\textbf {\bibinfo {volume} {94}},\ \bibinfo
  {pages} {013420} (\bibinfo {year} {2016})}\BibitemShut {NoStop}%
\bibitem [{\citenamefont {Ewald}\ \emph {et~al.}(2019)\citenamefont {Ewald},
  \citenamefont {Feldker}, \citenamefont {Hirzler}, \citenamefont {F{\"u}rst},\
  and\ \citenamefont {Gerritsma}}]{Ewald2019Observation}%
  \BibitemOpen
  \bibfield  {author} {\bibinfo {author} {\bibfnamefont {N.~V.}\ \bibnamefont
  {Ewald}}, \bibinfo {author} {\bibfnamefont {T.}~\bibnamefont {Feldker}},
  \bibinfo {author} {\bibfnamefont {H.}~\bibnamefont {Hirzler}}, \bibinfo
  {author} {\bibfnamefont {H.~A.}\ \bibnamefont {F{\"u}rst}}, \ and\ \bibinfo
  {author} {\bibfnamefont {R.}~\bibnamefont {Gerritsma}},\ }\href {\doibase
  10.1103/PhysRevLett.122.253401} {\bibfield  {journal} {\bibinfo  {journal}
  {Phys. Rev. Lett.}\ }\textbf {\bibinfo {volume} {122}},\ \bibinfo {pages}
  {253401} (\bibinfo {year} {2019})}\BibitemShut {NoStop}%
\bibitem [{\citenamefont {Secker}\ \emph {et~al.}(2017)\citenamefont {Secker},
  \citenamefont {Ewald}, \citenamefont {Joger}, \citenamefont {F{\"u}rst},
  \citenamefont {Feldker},\ and\ \citenamefont
  {Gerritsma}}]{Secker2017Trapped}%
  \BibitemOpen
  \bibfield  {author} {\bibinfo {author} {\bibfnamefont {T.}~\bibnamefont
  {Secker}}, \bibinfo {author} {\bibfnamefont {N.}~\bibnamefont {Ewald}},
  \bibinfo {author} {\bibfnamefont {J.}~\bibnamefont {Joger}}, \bibinfo
  {author} {\bibfnamefont {H.}~\bibnamefont {F{\"u}rst}}, \bibinfo {author}
  {\bibfnamefont {T.}~\bibnamefont {Feldker}}, \ and\ \bibinfo {author}
  {\bibfnamefont {R.}~\bibnamefont {Gerritsma}},\ }\href {\doibase
  10.1103/PhysRevLett.118.263201} {\bibfield  {journal} {\bibinfo  {journal}
  {Phys. Rev. Lett.}\ }\textbf {\bibinfo {volume} {118}},\ \bibinfo {pages}
  {263201} (\bibinfo {year} {2017})}\BibitemShut {NoStop}%
\bibitem [{\citenamefont {Schmid}\ \emph {et~al.}(2018)\citenamefont {Schmid},
  \citenamefont {Veit}, \citenamefont {Zuber}, \citenamefont {L{\"o}w},
  \citenamefont {Pfau}, \citenamefont {Tarana},\ and\ \citenamefont
  {Tomza}}]{Schmid2018Rydberg}%
  \BibitemOpen
  \bibfield  {author} {\bibinfo {author} {\bibfnamefont {T.}~\bibnamefont
  {Schmid}}, \bibinfo {author} {\bibfnamefont {C.}~\bibnamefont {Veit}},
  \bibinfo {author} {\bibfnamefont {N.}~\bibnamefont {Zuber}}, \bibinfo
  {author} {\bibfnamefont {R.}~\bibnamefont {L{\"o}w}}, \bibinfo {author}
  {\bibfnamefont {T.}~\bibnamefont {Pfau}}, \bibinfo {author} {\bibfnamefont
  {M.}~\bibnamefont {Tarana}}, \ and\ \bibinfo {author} {\bibfnamefont
  {M.}~\bibnamefont {Tomza}},\ }\href {\doibase 10.1103/PhysRevLett.120.153401}
  {\bibfield  {journal} {\bibinfo  {journal} {Phys. Rev. Lett.}\ }\textbf
  {\bibinfo {volume} {120}},\ \bibinfo {pages} {153401} (\bibinfo {year}
  {2018})}\BibitemShut {NoStop}%
\bibitem [{\citenamefont {C{\^o}t{\'e}}(2000)}]{Cote2000Classical}%
  \BibitemOpen
  \bibfield  {author} {\bibinfo {author} {\bibfnamefont {R.}~\bibnamefont
  {C{\^o}t{\'e}}},\ }\href {\doibase 10.1103/PhysRevLett.85.5316} {\bibfield
  {journal} {\bibinfo  {journal} {Phys. Rev. Lett.}\ }\textbf {\bibinfo
  {volume} {85}},\ \bibinfo {pages} {5316} (\bibinfo {year}
  {2000})}\BibitemShut {NoStop}%
\bibitem [{\citenamefont {Dieterle}\ \emph {et~al.}(2021)\citenamefont
  {Dieterle}, \citenamefont {Berngruber}, \citenamefont {H{\"o}lzl},
  \citenamefont {L{\"o}w}, \citenamefont {Jachymski}, \citenamefont {Pfau},\
  and\ \citenamefont {Meinert}}]{Dieterle2021Transport}%
  \BibitemOpen
  \bibfield  {author} {\bibinfo {author} {\bibfnamefont {T.}~\bibnamefont
  {Dieterle}}, \bibinfo {author} {\bibfnamefont {M.}~\bibnamefont
  {Berngruber}}, \bibinfo {author} {\bibfnamefont {C.}~\bibnamefont
  {H{\"o}lzl}}, \bibinfo {author} {\bibfnamefont {R.}~\bibnamefont {L{\"o}w}},
  \bibinfo {author} {\bibfnamefont {K.}~\bibnamefont {Jachymski}}, \bibinfo
  {author} {\bibfnamefont {T.}~\bibnamefont {Pfau}}, \ and\ \bibinfo {author}
  {\bibfnamefont {F.}~\bibnamefont {Meinert}},\ }\href {\doibase
  10.1103/PhysRevLett.126.033401} {\bibfield  {journal} {\bibinfo  {journal}
  {Phys. Rev. Lett.}\ }\textbf {\bibinfo {volume} {126}},\ \bibinfo {pages}
  {033401} (\bibinfo {year} {2021})}\BibitemShut {NoStop}%
\bibitem [{\citenamefont {Engel}\ \emph {et~al.}(2018)\citenamefont {Engel},
  \citenamefont {Dieterle}, \citenamefont {Schmid}, \citenamefont {Tomschitz},
  \citenamefont {Veit}, \citenamefont {Zuber}, \citenamefont {L{\"o}w},
  \citenamefont {Pfau},\ and\ \citenamefont {Meinert}}]{Engel2018Observation}%
  \BibitemOpen
  \bibfield  {author} {\bibinfo {author} {\bibfnamefont {F.}~\bibnamefont
  {Engel}}, \bibinfo {author} {\bibfnamefont {T.}~\bibnamefont {Dieterle}},
  \bibinfo {author} {\bibfnamefont {T.}~\bibnamefont {Schmid}}, \bibinfo
  {author} {\bibfnamefont {C.}~\bibnamefont {Tomschitz}}, \bibinfo {author}
  {\bibfnamefont {C.}~\bibnamefont {Veit}}, \bibinfo {author} {\bibfnamefont
  {N.}~\bibnamefont {Zuber}}, \bibinfo {author} {\bibfnamefont
  {R.}~\bibnamefont {L{\"o}w}}, \bibinfo {author} {\bibfnamefont
  {T.}~\bibnamefont {Pfau}}, \ and\ \bibinfo {author} {\bibfnamefont
  {F.}~\bibnamefont {Meinert}},\ }\href {\doibase
  10.1103/PhysRevLett.121.193401} {\bibfield  {journal} {\bibinfo  {journal}
  {Phys. Rev. Lett.}\ }\textbf {\bibinfo {volume} {121}},\ \bibinfo {pages}
  {193401} (\bibinfo {year} {2018})}\BibitemShut {NoStop}%
\bibitem [{\citenamefont {Kleinbach}\ \emph {et~al.}(2018)\citenamefont
  {Kleinbach}, \citenamefont {Engel}, \citenamefont {Dieterle}, \citenamefont
  {L{\"o}w}, \citenamefont {Pfau},\ and\ \citenamefont
  {Meinert}}]{Kleinbach2018Ionic}%
  \BibitemOpen
  \bibfield  {author} {\bibinfo {author} {\bibfnamefont {K.~S.}\ \bibnamefont
  {Kleinbach}}, \bibinfo {author} {\bibfnamefont {F.}~\bibnamefont {Engel}},
  \bibinfo {author} {\bibfnamefont {T.}~\bibnamefont {Dieterle}}, \bibinfo
  {author} {\bibfnamefont {R.}~\bibnamefont {L{\"o}w}}, \bibinfo {author}
  {\bibfnamefont {T.}~\bibnamefont {Pfau}}, \ and\ \bibinfo {author}
  {\bibfnamefont {F.}~\bibnamefont {Meinert}},\ }\href {\doibase
  10.1103/PhysRevLett.120.193401} {\bibfield  {journal} {\bibinfo  {journal}
  {Phys. Rev. Lett.}\ }\textbf {\bibinfo {volume} {120}},\ \bibinfo {pages}
  {193401} (\bibinfo {year} {2018})}\BibitemShut {NoStop}%
\bibitem [{\citenamefont {Duspayev}\ \emph {et~al.}(2021)\citenamefont
  {Duspayev}, \citenamefont {Han}, \citenamefont {Viray}, \citenamefont {Ma},
  \citenamefont {Zhao},\ and\ \citenamefont {Raithel}}]{Duspayev2021Longrange}%
  \BibitemOpen
  \bibfield  {author} {\bibinfo {author} {\bibfnamefont {A.}~\bibnamefont
  {Duspayev}}, \bibinfo {author} {\bibfnamefont {X.}~\bibnamefont {Han}},
  \bibinfo {author} {\bibfnamefont {M.~A.}\ \bibnamefont {Viray}}, \bibinfo
  {author} {\bibfnamefont {L.}~\bibnamefont {Ma}}, \bibinfo {author}
  {\bibfnamefont {J.}~\bibnamefont {Zhao}}, \ and\ \bibinfo {author}
  {\bibfnamefont {G.}~\bibnamefont {Raithel}},\ }\href {\doibase
  10.1103/PhysRevResearch.3.023114} {\bibfield  {journal} {\bibinfo  {journal}
  {Phys. Rev. Research}\ }\textbf {\bibinfo {volume} {3}},\ \bibinfo {pages}
  {023114} (\bibinfo {year} {2021})}\BibitemShut {NoStop}%
\bibitem [{\citenamefont {Dei{\ss}}\ \emph {et~al.}(2021)\citenamefont
  {Dei{\ss}}, \citenamefont {Haze},\ and\ \citenamefont
  {Hecker~Denschlag}}]{Deiss2021LongRange}%
  \BibitemOpen
  \bibfield  {author} {\bibinfo {author} {\bibfnamefont {M.}~\bibnamefont
  {Dei{\ss}}}, \bibinfo {author} {\bibfnamefont {S.}~\bibnamefont {Haze}}, \
  and\ \bibinfo {author} {\bibfnamefont {J.}~\bibnamefont {Hecker~Denschlag}},\
  }\href {\doibase 10.3390/atoms9020034} {\bibfield  {journal} {\bibinfo
  {journal} {Atoms}\ }\textbf {\bibinfo {volume} {9}},\ \bibinfo {pages} {34}
  (\bibinfo {year} {2021})}\BibitemShut {NoStop}%
\bibitem [{\citenamefont {Zuber}\ \emph {et~al.}(2022)\citenamefont {Zuber},
  \citenamefont {Anasuri}, \citenamefont {Berngruber}, \citenamefont {Zou},
  \citenamefont {Meinert}, \citenamefont {L{\"o}w},\ and\ \citenamefont
  {Pfau}}]{Zuber2022Observation}%
  \BibitemOpen
  \bibfield  {author} {\bibinfo {author} {\bibfnamefont {N.}~\bibnamefont
  {Zuber}}, \bibinfo {author} {\bibfnamefont {V.~S.~V.}\ \bibnamefont
  {Anasuri}}, \bibinfo {author} {\bibfnamefont {M.}~\bibnamefont {Berngruber}},
  \bibinfo {author} {\bibfnamefont {Y.-Q.}\ \bibnamefont {Zou}}, \bibinfo
  {author} {\bibfnamefont {F.}~\bibnamefont {Meinert}}, \bibinfo {author}
  {\bibfnamefont {R.}~\bibnamefont {L{\"o}w}}, \ and\ \bibinfo {author}
  {\bibfnamefont {T.}~\bibnamefont {Pfau}},\ }\href {\doibase
  10.1038/s41586-022-04577-5} {\bibfield  {journal} {\bibinfo  {journal}
  {Nature}\ }\textbf {\bibinfo {volume} {605}},\ \bibinfo {pages} {453}
  (\bibinfo {year} {2022})}\BibitemShut {NoStop}%
\bibitem [{\citenamefont {Zou}\ \emph {et~al.}(2022)\citenamefont {Zou},
  \citenamefont {Berngruber}, \citenamefont {Anasuri}, \citenamefont {Zuber},
  \citenamefont {Meinert}, \citenamefont {L{\"o}w},\ and\ \citenamefont
  {Pfau}}]{Zou2022Observation}%
  \BibitemOpen
  \bibfield  {author} {\bibinfo {author} {\bibfnamefont {Y.-Q.}\ \bibnamefont
  {Zou}}, \bibinfo {author} {\bibfnamefont {M.}~\bibnamefont {Berngruber}},
  \bibinfo {author} {\bibfnamefont {V.~S.~V.}\ \bibnamefont {Anasuri}},
  \bibinfo {author} {\bibfnamefont {N.}~\bibnamefont {Zuber}}, \bibinfo
  {author} {\bibfnamefont {F.}~\bibnamefont {Meinert}}, \bibinfo {author}
  {\bibfnamefont {R.}~\bibnamefont {L{\"o}w}}, \ and\ \bibinfo {author}
  {\bibfnamefont {T.}~\bibnamefont {Pfau}},\ }\href@noop {} {\  (\bibinfo
  {year} {2022})},\ \Eprint {http://arxiv.org/abs/2208.07776}
  {arXiv:2208.07776} \BibitemShut {NoStop}%
\bibitem [{\citenamefont {Eiles}(2019)}]{Eiles2019Trilobites}%
  \BibitemOpen
  \bibfield  {author} {\bibinfo {author} {\bibfnamefont {M.~T.}\ \bibnamefont
  {Eiles}},\ }\href {\doibase 10.1088/1361-6455/ab19ca} {\bibfield  {journal}
  {\bibinfo  {journal} {J. Phys. B: At. Mol. Opt. Phys.}\ }\textbf {\bibinfo
  {volume} {52}},\ \bibinfo {pages} {113001} (\bibinfo {year}
  {2019})}\BibitemShut {NoStop}%
\bibitem [{\citenamefont {Fey}\ \emph {et~al.}(2020)\citenamefont {Fey},
  \citenamefont {Hummel},\ and\ \citenamefont
  {Schmelcher}}]{Fey2020Ultralongrange}%
  \BibitemOpen
  \bibfield  {author} {\bibinfo {author} {\bibfnamefont {C.}~\bibnamefont
  {Fey}}, \bibinfo {author} {\bibfnamefont {F.}~\bibnamefont {Hummel}}, \ and\
  \bibinfo {author} {\bibfnamefont {P.}~\bibnamefont {Schmelcher}},\ }\href
  {\doibase 10.1080/00268976.2019.1679401} {\bibfield  {journal} {\bibinfo
  {journal} {Mol. Phys.}\ }\textbf {\bibinfo {volume} {118}},\ \bibinfo {pages}
  {e1679401} (\bibinfo {year} {2020})}\BibitemShut {NoStop}%
\bibitem [{\citenamefont {Fermi}(1934)}]{Fermi1934Sopra}%
  \BibitemOpen
  \bibfield  {author} {\bibinfo {author} {\bibfnamefont {E.}~\bibnamefont
  {Fermi}},\ }\href {\doibase 10.1007/BF02959829} {\bibfield  {journal}
  {\bibinfo  {journal} {Nuovo Cim}\ }\textbf {\bibinfo {volume} {11}},\
  \bibinfo {pages} {157} (\bibinfo {year} {1934})}\BibitemShut {NoStop}%
\bibitem [{\citenamefont {Greene}\ \emph {et~al.}(2000)\citenamefont {Greene},
  \citenamefont {Dickinson},\ and\ \citenamefont
  {Sadeghpour}}]{Greene2000Creation}%
  \BibitemOpen
  \bibfield  {author} {\bibinfo {author} {\bibfnamefont {C.~H.}\ \bibnamefont
  {Greene}}, \bibinfo {author} {\bibfnamefont {A.~S.}\ \bibnamefont
  {Dickinson}}, \ and\ \bibinfo {author} {\bibfnamefont {H.~R.}\ \bibnamefont
  {Sadeghpour}},\ }\href {\doibase 10.1103/PhysRevLett.85.2458} {\bibfield
  {journal} {\bibinfo  {journal} {Phys. Rev. Lett.}\ }\textbf {\bibinfo
  {volume} {85}},\ \bibinfo {pages} {2458} (\bibinfo {year}
  {2000})}\BibitemShut {NoStop}%
\bibitem [{\citenamefont {Bendkowsky}\ \emph {et~al.}(2009)\citenamefont
  {Bendkowsky}, \citenamefont {Butscher}, \citenamefont {Nipper}, \citenamefont
  {Shaffer}, \citenamefont {L{\"o}w},\ and\ \citenamefont
  {Pfau}}]{Bendkowsky2009Observation}%
  \BibitemOpen
  \bibfield  {author} {\bibinfo {author} {\bibfnamefont {V.}~\bibnamefont
  {Bendkowsky}}, \bibinfo {author} {\bibfnamefont {B.}~\bibnamefont
  {Butscher}}, \bibinfo {author} {\bibfnamefont {J.}~\bibnamefont {Nipper}},
  \bibinfo {author} {\bibfnamefont {J.~P.}\ \bibnamefont {Shaffer}}, \bibinfo
  {author} {\bibfnamefont {R.}~\bibnamefont {L{\"o}w}}, \ and\ \bibinfo
  {author} {\bibfnamefont {T.}~\bibnamefont {Pfau}},\ }\href {\doibase
  10.1038/nature07945} {\bibfield  {journal} {\bibinfo  {journal} {Nature}\
  }\textbf {\bibinfo {volume} {458}},\ \bibinfo {pages} {1005} (\bibinfo {year}
  {2009})}\BibitemShut {NoStop}%
\bibitem [{\citenamefont {Manthey}\ \emph {et~al.}(2015)\citenamefont
  {Manthey}, \citenamefont {Niederpr{\"u}m}, \citenamefont {Thomas},\ and\
  \citenamefont {Ott}}]{Manthey2015Dynamically}%
  \BibitemOpen
  \bibfield  {author} {\bibinfo {author} {\bibfnamefont {T.}~\bibnamefont
  {Manthey}}, \bibinfo {author} {\bibfnamefont {T.}~\bibnamefont
  {Niederpr{\"u}m}}, \bibinfo {author} {\bibfnamefont {O.}~\bibnamefont
  {Thomas}}, \ and\ \bibinfo {author} {\bibfnamefont {H.}~\bibnamefont {Ott}},\
  }\href {\doibase 10.1088/1367-2630/17/10/103024} {\bibfield  {journal}
  {\bibinfo  {journal} {New J. Phys.}\ }\textbf {\bibinfo {volume} {17}},\
  \bibinfo {pages} {103024} (\bibinfo {year} {2015})}\BibitemShut {NoStop}%
\bibitem [{\citenamefont {Whalen}\ \emph
  {et~al.}(2019{\natexlab{a}})\citenamefont {Whalen}, \citenamefont {Kanungo},
  \citenamefont {Ding}, \citenamefont {Wagner}, \citenamefont {Schmidt},
  \citenamefont {Sadeghpour}, \citenamefont {Yoshida}, \citenamefont
  {Burgd{\"o}rfer}, \citenamefont {Dunning},\ and\ \citenamefont
  {Killian}}]{Whalen2019Probing}%
  \BibitemOpen
  \bibfield  {author} {\bibinfo {author} {\bibfnamefont {J.~D.}\ \bibnamefont
  {Whalen}}, \bibinfo {author} {\bibfnamefont {S.~K.}\ \bibnamefont {Kanungo}},
  \bibinfo {author} {\bibfnamefont {R.}~\bibnamefont {Ding}}, \bibinfo {author}
  {\bibfnamefont {M.}~\bibnamefont {Wagner}}, \bibinfo {author} {\bibfnamefont
  {R.}~\bibnamefont {Schmidt}}, \bibinfo {author} {\bibfnamefont {H.~R.}\
  \bibnamefont {Sadeghpour}}, \bibinfo {author} {\bibfnamefont
  {S.}~\bibnamefont {Yoshida}}, \bibinfo {author} {\bibfnamefont
  {J.}~\bibnamefont {Burgd{\"o}rfer}}, \bibinfo {author} {\bibfnamefont
  {F.~B.}\ \bibnamefont {Dunning}}, \ and\ \bibinfo {author} {\bibfnamefont
  {T.~C.}\ \bibnamefont {Killian}},\ }\href {\doibase
  10.1103/PhysRevA.100.011402} {\bibfield  {journal} {\bibinfo  {journal}
  {Phys. Rev. A}\ }\textbf {\bibinfo {volume} {100}},\ \bibinfo {pages}
  {011402} (\bibinfo {year} {2019}{\natexlab{a}})}\BibitemShut {NoStop}%
\bibitem [{\citenamefont {Whalen}\ \emph
  {et~al.}(2019{\natexlab{b}})\citenamefont {Whalen}, \citenamefont {Ding},
  \citenamefont {Kanungo}, \citenamefont {Killian}, \citenamefont {Yoshida},
  \citenamefont {Burgd{\"o}rfer},\ and\ \citenamefont
  {Dunning}}]{Whalen2019Formation}%
  \BibitemOpen
  \bibfield  {author} {\bibinfo {author} {\bibfnamefont {J.~D.}\ \bibnamefont
  {Whalen}}, \bibinfo {author} {\bibfnamefont {R.}~\bibnamefont {Ding}},
  \bibinfo {author} {\bibfnamefont {S.~K.}\ \bibnamefont {Kanungo}}, \bibinfo
  {author} {\bibfnamefont {T.~C.}\ \bibnamefont {Killian}}, \bibinfo {author}
  {\bibfnamefont {S.}~\bibnamefont {Yoshida}}, \bibinfo {author} {\bibfnamefont
  {J.}~\bibnamefont {Burgd{\"o}rfer}}, \ and\ \bibinfo {author} {\bibfnamefont
  {F.~B.}\ \bibnamefont {Dunning}},\ }\href {\doibase
  10.1080/00268976.2019.1575485} {\bibfield  {journal} {\bibinfo  {journal}
  {Mol. Phys.}\ }\textbf {\bibinfo {volume} {117}},\ \bibinfo {pages} {3088}
  (\bibinfo {year} {2019}{\natexlab{b}})}\BibitemShut {NoStop}%
\bibitem [{\citenamefont {Anderson}\ \emph {et~al.}(2014)\citenamefont
  {Anderson}, \citenamefont {Miller},\ and\ \citenamefont
  {Raithel}}]{Anderson2014Photoassociation}%
  \BibitemOpen
  \bibfield  {author} {\bibinfo {author} {\bibfnamefont {D.~A.}\ \bibnamefont
  {Anderson}}, \bibinfo {author} {\bibfnamefont {S.~A.}\ \bibnamefont
  {Miller}}, \ and\ \bibinfo {author} {\bibfnamefont {G.}~\bibnamefont
  {Raithel}},\ }\href {\doibase 10.1103/PhysRevLett.112.163201} {\bibfield
  {journal} {\bibinfo  {journal} {Phys. Rev. Lett.}\ }\textbf {\bibinfo
  {volume} {112}},\ \bibinfo {pages} {163201} (\bibinfo {year}
  {2014})}\BibitemShut {NoStop}%
\bibitem [{\citenamefont {Sa{\ss}mannshausen}\ \emph
  {et~al.}(2015)\citenamefont {Sa{\ss}mannshausen}, \citenamefont {Merkt},\
  and\ \citenamefont {Deiglmayr}}]{Sassmannshausen2015Experimental}%
  \BibitemOpen
  \bibfield  {author} {\bibinfo {author} {\bibfnamefont {H.}~\bibnamefont
  {Sa{\ss}mannshausen}}, \bibinfo {author} {\bibfnamefont {F.}~\bibnamefont
  {Merkt}}, \ and\ \bibinfo {author} {\bibfnamefont {J.}~\bibnamefont
  {Deiglmayr}},\ }\href {\doibase 10.1103/PhysRevLett.114.133201} {\bibfield
  {journal} {\bibinfo  {journal} {Phys. Rev. Lett.}\ }\textbf {\bibinfo
  {volume} {114}},\ \bibinfo {pages} {133201} (\bibinfo {year}
  {2015})}\BibitemShut {NoStop}%
\bibitem [{\citenamefont {B{\"o}ttcher}\ \emph {et~al.}(2016)\citenamefont
  {B{\"o}ttcher}, \citenamefont {Gaj}, \citenamefont {Westphal}, \citenamefont
  {Schlagm{\"u}ller}, \citenamefont {Kleinbach}, \citenamefont {L{\"o}w},
  \citenamefont {Liebisch}, \citenamefont {Pfau},\ and\ \citenamefont
  {Hofferberth}}]{Bottcher2016Observation}%
  \BibitemOpen
  \bibfield  {author} {\bibinfo {author} {\bibfnamefont {F.}~\bibnamefont
  {B{\"o}ttcher}}, \bibinfo {author} {\bibfnamefont {A.}~\bibnamefont {Gaj}},
  \bibinfo {author} {\bibfnamefont {K.~M.}\ \bibnamefont {Westphal}}, \bibinfo
  {author} {\bibfnamefont {M.}~\bibnamefont {Schlagm{\"u}ller}}, \bibinfo
  {author} {\bibfnamefont {K.~S.}\ \bibnamefont {Kleinbach}}, \bibinfo {author}
  {\bibfnamefont {R.}~\bibnamefont {L{\"o}w}}, \bibinfo {author} {\bibfnamefont
  {T.~C.}\ \bibnamefont {Liebisch}}, \bibinfo {author} {\bibfnamefont
  {T.}~\bibnamefont {Pfau}}, \ and\ \bibinfo {author} {\bibfnamefont
  {S.}~\bibnamefont {Hofferberth}},\ }\href {\doibase
  10.1103/PhysRevA.93.032512} {\bibfield  {journal} {\bibinfo  {journal} {Phys.
  Rev. A}\ }\textbf {\bibinfo {volume} {93}},\ \bibinfo {pages} {032512}
  (\bibinfo {year} {2016})}\BibitemShut {NoStop}%
\bibitem [{\citenamefont {Schmidt}\ \emph {et~al.}(2016)\citenamefont
  {Schmidt}, \citenamefont {Sadeghpour},\ and\ \citenamefont
  {Demler}}]{Schmidt2016Mesoscopic}%
  \BibitemOpen
  \bibfield  {author} {\bibinfo {author} {\bibfnamefont {R.}~\bibnamefont
  {Schmidt}}, \bibinfo {author} {\bibfnamefont {H.~R.}\ \bibnamefont
  {Sadeghpour}}, \ and\ \bibinfo {author} {\bibfnamefont {E.}~\bibnamefont
  {Demler}},\ }\href {\doibase 10.1103/PhysRevLett.116.105302} {\bibfield
  {journal} {\bibinfo  {journal} {Phys. Rev. Lett.}\ }\textbf {\bibinfo
  {volume} {116}},\ \bibinfo {pages} {105302} (\bibinfo {year}
  {2016})}\BibitemShut {NoStop}%
\bibitem [{\citenamefont {Camargo}\ \emph {et~al.}(2018)\citenamefont
  {Camargo}, \citenamefont {Schmidt}, \citenamefont {Whalen}, \citenamefont
  {Ding}, \citenamefont {Woehl}, \citenamefont {Yoshida}, \citenamefont
  {Burgd{\"o}rfer}, \citenamefont {Dunning}, \citenamefont {Sadeghpour},
  \citenamefont {Demler},\ and\ \citenamefont {Killian}}]{Camargo2018Creation}%
  \BibitemOpen
  \bibfield  {author} {\bibinfo {author} {\bibfnamefont {F.}~\bibnamefont
  {Camargo}}, \bibinfo {author} {\bibfnamefont {R.}~\bibnamefont {Schmidt}},
  \bibinfo {author} {\bibfnamefont {J.~D.}\ \bibnamefont {Whalen}}, \bibinfo
  {author} {\bibfnamefont {R.}~\bibnamefont {Ding}}, \bibinfo {author}
  {\bibfnamefont {G.}~\bibnamefont {Woehl}}, \bibinfo {author} {\bibfnamefont
  {S.}~\bibnamefont {Yoshida}}, \bibinfo {author} {\bibfnamefont
  {J.}~\bibnamefont {Burgd{\"o}rfer}}, \bibinfo {author} {\bibfnamefont
  {F.~B.}\ \bibnamefont {Dunning}}, \bibinfo {author} {\bibfnamefont {H.~R.}\
  \bibnamefont {Sadeghpour}}, \bibinfo {author} {\bibfnamefont
  {E.}~\bibnamefont {Demler}}, \ and\ \bibinfo {author} {\bibfnamefont {T.~C.}\
  \bibnamefont {Killian}},\ }\href {\doibase 10.1103/PhysRevLett.120.083401}
  {\bibfield  {journal} {\bibinfo  {journal} {Phys. Rev. Lett.}\ }\textbf
  {\bibinfo {volume} {120}},\ \bibinfo {pages} {083401} (\bibinfo {year}
  {2018})}\BibitemShut {NoStop}%
\bibitem [{\citenamefont {Sous}\ \emph {et~al.}(2020)\citenamefont {Sous},
  \citenamefont {Sadeghpour}, \citenamefont {Killian}, \citenamefont {Demler},\
  and\ \citenamefont {Schmidt}}]{Sous2020Rydberg}%
  \BibitemOpen
  \bibfield  {author} {\bibinfo {author} {\bibfnamefont {J.}~\bibnamefont
  {Sous}}, \bibinfo {author} {\bibfnamefont {H.~R.}\ \bibnamefont
  {Sadeghpour}}, \bibinfo {author} {\bibfnamefont {T.~C.}\ \bibnamefont
  {Killian}}, \bibinfo {author} {\bibfnamefont {E.}~\bibnamefont {Demler}}, \
  and\ \bibinfo {author} {\bibfnamefont {R.}~\bibnamefont {Schmidt}},\ }\href
  {\doibase 10.1103/PhysRevResearch.2.023021} {\bibfield  {journal} {\bibinfo
  {journal} {Phys. Rev. Research}\ }\textbf {\bibinfo {volume} {2}},\ \bibinfo
  {pages} {023021} (\bibinfo {year} {2020})}\BibitemShut {NoStop}%
\bibitem [{\citenamefont {Bendkowsky}\ \emph {et~al.}(2010)\citenamefont
  {Bendkowsky}, \citenamefont {Butscher}, \citenamefont {Nipper}, \citenamefont
  {Balewski}, \citenamefont {Shaffer}, \citenamefont {L{\"o}w}, \citenamefont
  {Pfau}, \citenamefont {Li}, \citenamefont {Stanojevic}, \citenamefont
  {Pohl},\ and\ \citenamefont {Rost}}]{Bendkowsky2010Rydberg}%
  \BibitemOpen
  \bibfield  {author} {\bibinfo {author} {\bibfnamefont {V.}~\bibnamefont
  {Bendkowsky}}, \bibinfo {author} {\bibfnamefont {B.}~\bibnamefont
  {Butscher}}, \bibinfo {author} {\bibfnamefont {J.}~\bibnamefont {Nipper}},
  \bibinfo {author} {\bibfnamefont {J.~B.}\ \bibnamefont {Balewski}}, \bibinfo
  {author} {\bibfnamefont {J.~P.}\ \bibnamefont {Shaffer}}, \bibinfo {author}
  {\bibfnamefont {R.}~\bibnamefont {L{\"o}w}}, \bibinfo {author} {\bibfnamefont
  {T.}~\bibnamefont {Pfau}}, \bibinfo {author} {\bibfnamefont {W.}~\bibnamefont
  {Li}}, \bibinfo {author} {\bibfnamefont {J.}~\bibnamefont {Stanojevic}},
  \bibinfo {author} {\bibfnamefont {T.}~\bibnamefont {Pohl}}, \ and\ \bibinfo
  {author} {\bibfnamefont {J.~M.}\ \bibnamefont {Rost}},\ }\href {\doibase
  10.1103/PhysRevLett.105.163201} {\bibfield  {journal} {\bibinfo  {journal}
  {Phys. Rev. Lett.}\ }\textbf {\bibinfo {volume} {105}},\ \bibinfo {pages}
  {163201} (\bibinfo {year} {2010})}\BibitemShut {NoStop}%
\bibitem [{\citenamefont {Gaj}\ \emph {et~al.}(2014)\citenamefont {Gaj},
  \citenamefont {Krupp}, \citenamefont {Balewski}, \citenamefont {L{\"o}w},
  \citenamefont {Hofferberth},\ and\ \citenamefont {Pfau}}]{Gaj2014Molecular}%
  \BibitemOpen
  \bibfield  {author} {\bibinfo {author} {\bibfnamefont {A.}~\bibnamefont
  {Gaj}}, \bibinfo {author} {\bibfnamefont {A.~T.}\ \bibnamefont {Krupp}},
  \bibinfo {author} {\bibfnamefont {J.~B.}\ \bibnamefont {Balewski}}, \bibinfo
  {author} {\bibfnamefont {R.}~\bibnamefont {L{\"o}w}}, \bibinfo {author}
  {\bibfnamefont {S.}~\bibnamefont {Hofferberth}}, \ and\ \bibinfo {author}
  {\bibfnamefont {T.}~\bibnamefont {Pfau}},\ }\href {\doibase
  10.1038/ncomms5546} {\bibfield  {journal} {\bibinfo  {journal} {Nat.
  Commun.}\ }\textbf {\bibinfo {volume} {5}},\ \bibinfo {pages} {4546}
  (\bibinfo {year} {2014})}\BibitemShut {NoStop}%
\bibitem [{\citenamefont {Liu}\ \emph {et~al.}(2009)\citenamefont {Liu},
  \citenamefont {Stanojevic},\ and\ \citenamefont
  {Rost}}]{Liu2009UltraLongRange}%
  \BibitemOpen
  \bibfield  {author} {\bibinfo {author} {\bibfnamefont {I.~C.~H.}\
  \bibnamefont {Liu}}, \bibinfo {author} {\bibfnamefont {J.}~\bibnamefont
  {Stanojevic}}, \ and\ \bibinfo {author} {\bibfnamefont {J.~M.}\ \bibnamefont
  {Rost}},\ }\href {\doibase 10.1103/PhysRevLett.102.173001} {\bibfield
  {journal} {\bibinfo  {journal} {Phys. Rev. Lett.}\ }\textbf {\bibinfo
  {volume} {102}},\ \bibinfo {pages} {173001} (\bibinfo {year}
  {2009})}\BibitemShut {NoStop}%
\bibitem [{\citenamefont {Fey}\ \emph {et~al.}(2016)\citenamefont {Fey},
  \citenamefont {Kurz},\ and\ \citenamefont {Schmelcher}}]{Fey2016Stretching}%
  \BibitemOpen
  \bibfield  {author} {\bibinfo {author} {\bibfnamefont {C.}~\bibnamefont
  {Fey}}, \bibinfo {author} {\bibfnamefont {M.}~\bibnamefont {Kurz}}, \ and\
  \bibinfo {author} {\bibfnamefont {P.}~\bibnamefont {Schmelcher}},\ }\href
  {\doibase 10.1103/PhysRevA.94.012516} {\bibfield  {journal} {\bibinfo
  {journal} {Phys. Rev. A}\ }\textbf {\bibinfo {volume} {94}},\ \bibinfo
  {pages} {012516} (\bibinfo {year} {2016})}\BibitemShut {NoStop}%
\bibitem [{\citenamefont {Eiles}\ \emph {et~al.}(2016)\citenamefont {Eiles},
  \citenamefont {{P{\'e}rez-R{\'i}os}}, \citenamefont {Robicheaux},\ and\
  \citenamefont {Greene}}]{Eiles2016Ultracold}%
  \BibitemOpen
  \bibfield  {author} {\bibinfo {author} {\bibfnamefont {M.~T.}\ \bibnamefont
  {Eiles}}, \bibinfo {author} {\bibfnamefont {J.}~\bibnamefont
  {{P{\'e}rez-R{\'i}os}}}, \bibinfo {author} {\bibfnamefont {F.}~\bibnamefont
  {Robicheaux}}, \ and\ \bibinfo {author} {\bibfnamefont {C.~H.}\ \bibnamefont
  {Greene}},\ }\href {\doibase 10.1088/0953-4075/49/11/114005} {\bibfield
  {journal} {\bibinfo  {journal} {J. Phys. B: At. Mol. Opt. Phys.}\ }\textbf
  {\bibinfo {volume} {49}},\ \bibinfo {pages} {114005} (\bibinfo {year}
  {2016})}\BibitemShut {NoStop}%
\bibitem [{\citenamefont {Fey}\ \emph {et~al.}(2019{\natexlab{a}})\citenamefont
  {Fey}, \citenamefont {Yang}, \citenamefont {Rittenhouse}, \citenamefont
  {Munkes}, \citenamefont {Baluktsian}, \citenamefont {Schmelcher},
  \citenamefont {Sadeghpour},\ and\ \citenamefont
  {Shaffer}}]{Fey2019Effective}%
  \BibitemOpen
  \bibfield  {author} {\bibinfo {author} {\bibfnamefont {C.}~\bibnamefont
  {Fey}}, \bibinfo {author} {\bibfnamefont {J.}~\bibnamefont {Yang}}, \bibinfo
  {author} {\bibfnamefont {S.~T.}\ \bibnamefont {Rittenhouse}}, \bibinfo
  {author} {\bibfnamefont {F.}~\bibnamefont {Munkes}}, \bibinfo {author}
  {\bibfnamefont {M.}~\bibnamefont {Baluktsian}}, \bibinfo {author}
  {\bibfnamefont {P.}~\bibnamefont {Schmelcher}}, \bibinfo {author}
  {\bibfnamefont {H.~R.}\ \bibnamefont {Sadeghpour}}, \ and\ \bibinfo {author}
  {\bibfnamefont {J.~P.}\ \bibnamefont {Shaffer}},\ }\href {\doibase
  10.1103/PhysRevLett.122.103001} {\bibfield  {journal} {\bibinfo  {journal}
  {Phys. Rev. Lett.}\ }\textbf {\bibinfo {volume} {122}},\ \bibinfo {pages}
  {103001} (\bibinfo {year} {2019}{\natexlab{a}})}\BibitemShut {NoStop}%
\bibitem [{\citenamefont {Eiles}\ \emph {et~al.}(2020)\citenamefont {Eiles},
  \citenamefont {Fey}, \citenamefont {Hummel},\ and\ \citenamefont
  {Schmelcher}}]{Eiles2020Triatomic}%
  \BibitemOpen
  \bibfield  {author} {\bibinfo {author} {\bibfnamefont {M.~T.}\ \bibnamefont
  {Eiles}}, \bibinfo {author} {\bibfnamefont {C.}~\bibnamefont {Fey}}, \bibinfo
  {author} {\bibfnamefont {F.}~\bibnamefont {Hummel}}, \ and\ \bibinfo {author}
  {\bibfnamefont {P.}~\bibnamefont {Schmelcher}},\ }\href {\doibase
  10.1088/1361-6455/ab73af} {\bibfield  {journal} {\bibinfo  {journal} {J.
  Phys. B: At. Mol. Opt. Phys.}\ }\textbf {\bibinfo {volume} {53}},\ \bibinfo
  {pages} {084001} (\bibinfo {year} {2020})}\BibitemShut {NoStop}%
\bibitem [{\citenamefont {Hunter}\ \emph {et~al.}(2020)\citenamefont {Hunter},
  \citenamefont {Eiles}, \citenamefont {Eisfeld},\ and\ \citenamefont
  {Rost}}]{Hunter2020Rydberg}%
  \BibitemOpen
  \bibfield  {author} {\bibinfo {author} {\bibfnamefont {A.~L.}\ \bibnamefont
  {Hunter}}, \bibinfo {author} {\bibfnamefont {M.~T.}\ \bibnamefont {Eiles}},
  \bibinfo {author} {\bibfnamefont {A.}~\bibnamefont {Eisfeld}}, \ and\
  \bibinfo {author} {\bibfnamefont {J.~M.}\ \bibnamefont {Rost}},\ }\href
  {\doibase 10.1103/PhysRevX.10.031046} {\bibfield  {journal} {\bibinfo
  {journal} {Phys. Rev. X}\ }\textbf {\bibinfo {volume} {10}},\ \bibinfo
  {pages} {031046} (\bibinfo {year} {2020})}\BibitemShut {NoStop}%
\bibitem [{\citenamefont {Lesanovsky}\ \emph {et~al.}(2006)\citenamefont
  {Lesanovsky}, \citenamefont {Schmelcher},\ and\ \citenamefont
  {Sadeghpour}}]{Lesanovsky2006Ultralongrange}%
  \BibitemOpen
  \bibfield  {author} {\bibinfo {author} {\bibfnamefont {I.}~\bibnamefont
  {Lesanovsky}}, \bibinfo {author} {\bibfnamefont {P.}~\bibnamefont
  {Schmelcher}}, \ and\ \bibinfo {author} {\bibfnamefont {H.~R.}\ \bibnamefont
  {Sadeghpour}},\ }\href {\doibase 10.1088/0953-4075/39/4/L03} {\bibfield
  {journal} {\bibinfo  {journal} {J. Phys. B: At. Mol. Opt. Phys.}\ }\textbf
  {\bibinfo {volume} {39}},\ \bibinfo {pages} {L69} (\bibinfo {year}
  {2006})}\BibitemShut {NoStop}%
\bibitem [{\citenamefont {Kurz}\ and\ \citenamefont
  {Schmelcher}(2014)}]{Kurz2014Ultralongrange}%
  \BibitemOpen
  \bibfield  {author} {\bibinfo {author} {\bibfnamefont {M.}~\bibnamefont
  {Kurz}}\ and\ \bibinfo {author} {\bibfnamefont {P.}~\bibnamefont
  {Schmelcher}},\ }\href {\doibase 10.1088/0953-4075/47/16/165101} {\bibfield
  {journal} {\bibinfo  {journal} {J. Phys. B: At. Mol. Opt. Phys.}\ }\textbf
  {\bibinfo {volume} {47}},\ \bibinfo {pages} {165101} (\bibinfo {year}
  {2014})}\BibitemShut {NoStop}%
\bibitem [{\citenamefont {Krupp}\ \emph {et~al.}(2014)\citenamefont {Krupp},
  \citenamefont {Gaj}, \citenamefont {Balewski}, \citenamefont {Ilzh{\"o}fer},
  \citenamefont {Hofferberth}, \citenamefont {L{\"o}w}, \citenamefont {Pfau},
  \citenamefont {Kurz},\ and\ \citenamefont {Schmelcher}}]{Krupp2014Alignment}%
  \BibitemOpen
  \bibfield  {author} {\bibinfo {author} {\bibfnamefont {A.~T.}\ \bibnamefont
  {Krupp}}, \bibinfo {author} {\bibfnamefont {A.}~\bibnamefont {Gaj}}, \bibinfo
  {author} {\bibfnamefont {J.~B.}\ \bibnamefont {Balewski}}, \bibinfo {author}
  {\bibfnamefont {P.}~\bibnamefont {Ilzh{\"o}fer}}, \bibinfo {author}
  {\bibfnamefont {S.}~\bibnamefont {Hofferberth}}, \bibinfo {author}
  {\bibfnamefont {R.}~\bibnamefont {L{\"o}w}}, \bibinfo {author} {\bibfnamefont
  {T.}~\bibnamefont {Pfau}}, \bibinfo {author} {\bibfnamefont {M.}~\bibnamefont
  {Kurz}}, \ and\ \bibinfo {author} {\bibfnamefont {P.}~\bibnamefont
  {Schmelcher}},\ }\href {\doibase 10.1103/PhysRevLett.112.143008} {\bibfield
  {journal} {\bibinfo  {journal} {Phys. Rev. Lett.}\ }\textbf {\bibinfo
  {volume} {112}},\ \bibinfo {pages} {143008} (\bibinfo {year}
  {2014})}\BibitemShut {NoStop}%
\bibitem [{\citenamefont {Gaj}\ \emph {et~al.}(2015)\citenamefont {Gaj},
  \citenamefont {Krupp}, \citenamefont {Ilzh{\"o}fer}, \citenamefont {L{\"o}w},
  \citenamefont {Hofferberth},\ and\ \citenamefont
  {Pfau}}]{Gaj2015Hybridization}%
  \BibitemOpen
  \bibfield  {author} {\bibinfo {author} {\bibfnamefont {A.}~\bibnamefont
  {Gaj}}, \bibinfo {author} {\bibfnamefont {A.~T.}\ \bibnamefont {Krupp}},
  \bibinfo {author} {\bibfnamefont {P.}~\bibnamefont {Ilzh{\"o}fer}}, \bibinfo
  {author} {\bibfnamefont {R.}~\bibnamefont {L{\"o}w}}, \bibinfo {author}
  {\bibfnamefont {S.}~\bibnamefont {Hofferberth}}, \ and\ \bibinfo {author}
  {\bibfnamefont {T.}~\bibnamefont {Pfau}},\ }\href {\doibase
  10.1103/PhysRevLett.115.023001} {\bibfield  {journal} {\bibinfo  {journal}
  {Phys. Rev. Lett.}\ }\textbf {\bibinfo {volume} {115}},\ \bibinfo {pages}
  {023001} (\bibinfo {year} {2015})}\BibitemShut {NoStop}%
\bibitem [{\citenamefont {Niederpr{\"u}m}\ \emph {et~al.}(2016)\citenamefont
  {Niederpr{\"u}m}, \citenamefont {Thomas}, \citenamefont {Eichert},
  \citenamefont {Lippe}, \citenamefont {{P{\'e}rez-R{\'i}os}}, \citenamefont
  {Greene},\ and\ \citenamefont {Ott}}]{Niederprum2016Observation}%
  \BibitemOpen
  \bibfield  {author} {\bibinfo {author} {\bibfnamefont {T.}~\bibnamefont
  {Niederpr{\"u}m}}, \bibinfo {author} {\bibfnamefont {O.}~\bibnamefont
  {Thomas}}, \bibinfo {author} {\bibfnamefont {T.}~\bibnamefont {Eichert}},
  \bibinfo {author} {\bibfnamefont {C.}~\bibnamefont {Lippe}}, \bibinfo
  {author} {\bibfnamefont {J.}~\bibnamefont {{P{\'e}rez-R{\'i}os}}}, \bibinfo
  {author} {\bibfnamefont {C.~H.}\ \bibnamefont {Greene}}, \ and\ \bibinfo
  {author} {\bibfnamefont {H.}~\bibnamefont {Ott}},\ }\href {\doibase
  10.1038/ncomms12820} {\bibfield  {journal} {\bibinfo  {journal} {Nat.
  Commun.}\ }\textbf {\bibinfo {volume} {7}},\ \bibinfo {pages} {12820}
  (\bibinfo {year} {2016})}\BibitemShut {NoStop}%
\bibitem [{\citenamefont {Hummel}\ \emph {et~al.}(2019)\citenamefont {Hummel},
  \citenamefont {Fey},\ and\ \citenamefont {Schmelcher}}]{Hummel2019Alignment}%
  \BibitemOpen
  \bibfield  {author} {\bibinfo {author} {\bibfnamefont {F.}~\bibnamefont
  {Hummel}}, \bibinfo {author} {\bibfnamefont {C.}~\bibnamefont {Fey}}, \ and\
  \bibinfo {author} {\bibfnamefont {P.}~\bibnamefont {Schmelcher}},\ }\href
  {\doibase 10.1103/PhysRevA.99.023401} {\bibfield  {journal} {\bibinfo
  {journal} {Phys. Rev. A}\ }\textbf {\bibinfo {volume} {99}},\ \bibinfo
  {pages} {023401} (\bibinfo {year} {2019})}\BibitemShut {NoStop}%
\bibitem [{\citenamefont {Engel}\ \emph {et~al.}(2019)\citenamefont {Engel},
  \citenamefont {Dieterle}, \citenamefont {Hummel}, \citenamefont {Fey},
  \citenamefont {Schmelcher}, \citenamefont {L{\"o}w}, \citenamefont {Pfau},\
  and\ \citenamefont {Meinert}}]{Engel2019Precision}%
  \BibitemOpen
  \bibfield  {author} {\bibinfo {author} {\bibfnamefont {F.}~\bibnamefont
  {Engel}}, \bibinfo {author} {\bibfnamefont {T.}~\bibnamefont {Dieterle}},
  \bibinfo {author} {\bibfnamefont {F.}~\bibnamefont {Hummel}}, \bibinfo
  {author} {\bibfnamefont {C.}~\bibnamefont {Fey}}, \bibinfo {author}
  {\bibfnamefont {P.}~\bibnamefont {Schmelcher}}, \bibinfo {author}
  {\bibfnamefont {R.}~\bibnamefont {L{\"o}w}}, \bibinfo {author} {\bibfnamefont
  {T.}~\bibnamefont {Pfau}}, \ and\ \bibinfo {author} {\bibfnamefont
  {F.}~\bibnamefont {Meinert}},\ }\href {\doibase
  10.1103/PhysRevLett.123.073003} {\bibfield  {journal} {\bibinfo  {journal}
  {Phys. Rev. Lett.}\ }\textbf {\bibinfo {volume} {123}},\ \bibinfo {pages}
  {073003} (\bibinfo {year} {2019})}\BibitemShut {NoStop}%
\bibitem [{\citenamefont {Hummel}\ \emph
  {et~al.}(2021{\natexlab{a}})\citenamefont {Hummel}, \citenamefont {Keiler},\
  and\ \citenamefont {Schmelcher}}]{Hummel2021Electricfieldinduced}%
  \BibitemOpen
  \bibfield  {author} {\bibinfo {author} {\bibfnamefont {F.}~\bibnamefont
  {Hummel}}, \bibinfo {author} {\bibfnamefont {K.}~\bibnamefont {Keiler}}, \
  and\ \bibinfo {author} {\bibfnamefont {P.}~\bibnamefont {Schmelcher}},\
  }\href {\doibase 10.1103/PhysRevA.103.022827} {\bibfield  {journal} {\bibinfo
   {journal} {Phys. Rev. A}\ }\textbf {\bibinfo {volume} {103}},\ \bibinfo
  {pages} {022827} (\bibinfo {year} {2021}{\natexlab{a}})}\BibitemShut
  {NoStop}%
\bibitem [{\citenamefont {Hollerith}\ and\ \citenamefont
  {Zeiher}(2022)}]{Hollerith2022Rydberg}%
  \BibitemOpen
  \bibfield  {author} {\bibinfo {author} {\bibfnamefont {S.}~\bibnamefont
  {Hollerith}}\ and\ \bibinfo {author} {\bibfnamefont {J.}~\bibnamefont
  {Zeiher}},\ }\href@noop {} {\  (\bibinfo {year} {2022})},\ \Eprint
  {http://arxiv.org/abs/2212.01673} {arXiv:2212.01673} \BibitemShut {NoStop}%
\bibitem [{\citenamefont {Le~Roy}(1974)}]{LeRoy1974LongRange}%
  \BibitemOpen
  \bibfield  {author} {\bibinfo {author} {\bibfnamefont {R.~J.}\ \bibnamefont
  {Le~Roy}},\ }\href {\doibase 10.1139/p74-035} {\bibfield  {journal} {\bibinfo
   {journal} {Can. J. Phys.}\ }\textbf {\bibinfo {volume} {52}},\ \bibinfo
  {pages} {246} (\bibinfo {year} {1974})}\BibitemShut {NoStop}%
\bibitem [{\citenamefont {Duspayev}\ and\ \citenamefont
  {Raithel}(2022)}]{Duspayev2022Nonadiabatic}%
  \BibitemOpen
  \bibfield  {author} {\bibinfo {author} {\bibfnamefont {A.}~\bibnamefont
  {Duspayev}}\ and\ \bibinfo {author} {\bibfnamefont {G.}~\bibnamefont
  {Raithel}},\ }\href {\doibase 10.1103/PhysRevA.105.012810} {\bibfield
  {journal} {\bibinfo  {journal} {Phys. Rev. A}\ }\textbf {\bibinfo {volume}
  {105}},\ \bibinfo {pages} {012810} (\bibinfo {year} {2022})}\BibitemShut
  {NoStop}%
\bibitem [{\citenamefont {Fey}\ \emph {et~al.}(2019{\natexlab{b}})\citenamefont
  {Fey}, \citenamefont {Hummel},\ and\ \citenamefont
  {Schmelcher}}]{Fey2019Building}%
  \BibitemOpen
  \bibfield  {author} {\bibinfo {author} {\bibfnamefont {C.}~\bibnamefont
  {Fey}}, \bibinfo {author} {\bibfnamefont {F.}~\bibnamefont {Hummel}}, \ and\
  \bibinfo {author} {\bibfnamefont {P.}~\bibnamefont {Schmelcher}},\ }\href
  {\doibase 10.1103/PhysRevA.99.022506} {\bibfield  {journal} {\bibinfo
  {journal} {Phys. Rev. A}\ }\textbf {\bibinfo {volume} {99}},\ \bibinfo
  {pages} {022506} (\bibinfo {year} {2019}{\natexlab{b}})}\BibitemShut
  {NoStop}%
\bibitem [{\citenamefont {Carter}\ and\ \citenamefont
  {Handy}(1982)}]{Carter1982Variational}%
  \BibitemOpen
  \bibfield  {author} {\bibinfo {author} {\bibfnamefont {S.}~\bibnamefont
  {Carter}}\ and\ \bibinfo {author} {\bibfnamefont {N.}~\bibnamefont {Handy}},\
  }\href {\doibase 10.1080/00268978200101082} {\bibfield  {journal} {\bibinfo
  {journal} {Mol. Phys.}\ }\textbf {\bibinfo {volume} {47}},\ \bibinfo {pages}
  {1445} (\bibinfo {year} {1982})}\BibitemShut {NoStop}%
\bibitem [{\citenamefont {Handy}(1987)}]{Handy1987Derivation}%
  \BibitemOpen
  \bibfield  {author} {\bibinfo {author} {\bibfnamefont {N.}~\bibnamefont
  {Handy}},\ }\href {\doibase 10.1080/00268978700101081} {\bibfield  {journal}
  {\bibinfo  {journal} {Mol. Phys.}\ }\textbf {\bibinfo {volume} {61}},\
  \bibinfo {pages} {207} (\bibinfo {year} {1987})}\BibitemShut {NoStop}%
\bibitem [{\citenamefont {Meyer}\ \emph {et~al.}(1990)\citenamefont {Meyer},
  \citenamefont {Manthe},\ and\ \citenamefont {Cederbaum}}]{mey90:73}%
  \BibitemOpen
  \bibfield  {author} {\bibinfo {author} {\bibfnamefont {H.-D.}\ \bibnamefont
  {Meyer}}, \bibinfo {author} {\bibfnamefont {U.}~\bibnamefont {Manthe}}, \
  and\ \bibinfo {author} {\bibfnamefont {L.~S.}\ \bibnamefont {Cederbaum}},\
  }\href@noop {} {\bibfield  {journal} {\bibinfo  {journal} {Chem.\ Phys.\
  Lett.}\ }\textbf {\bibinfo {volume} {165}},\ \bibinfo {pages} {73} (\bibinfo
  {year} {1990})}\BibitemShut {NoStop}%
\bibitem [{\citenamefont {Beck}\ \emph {et~al.}(2000)\citenamefont {Beck},
  \citenamefont {J{\"a}ckle}, \citenamefont {Worth},\ and\ \citenamefont
  {Meyer}}]{bec00:1}%
  \BibitemOpen
  \bibfield  {author} {\bibinfo {author} {\bibfnamefont {M.~H.}\ \bibnamefont
  {Beck}}, \bibinfo {author} {\bibfnamefont {A.}~\bibnamefont {J{\"a}ckle}},
  \bibinfo {author} {\bibfnamefont {G.~A.}\ \bibnamefont {Worth}}, \ and\
  \bibinfo {author} {\bibfnamefont {H.-D.}\ \bibnamefont {Meyer}},\ }\href@noop
  {} {\bibfield  {journal} {\bibinfo  {journal} {Phys.\ Rep.}\ }\textbf
  {\bibinfo {volume} {324}},\ \bibinfo {pages} {1} (\bibinfo {year}
  {2000})}\BibitemShut {NoStop}%
\bibitem [{\citenamefont {Meyer}\ and\ \citenamefont
  {Worth}(2003)}]{mey03:251}%
  \BibitemOpen
  \bibfield  {author} {\bibinfo {author} {\bibfnamefont {H.-D.}\ \bibnamefont
  {Meyer}}\ and\ \bibinfo {author} {\bibfnamefont {G.~A.}\ \bibnamefont
  {Worth}},\ }\href@noop {} {\bibfield  {journal} {\bibinfo  {journal} {Theor.\
  Chem.\ Acc.}\ }\textbf {\bibinfo {volume} {109}},\ \bibinfo {pages} {251}
  (\bibinfo {year} {2003})}\BibitemShut {NoStop}%
\bibitem [{\citenamefont {Meyer}\ \emph {et~al.}(2006)\citenamefont {Meyer},
  \citenamefont {{Le Qu\'er\'e}}, \citenamefont {L\'eonard},\ and\
  \citenamefont {Gatti}}]{mey06:179}%
  \BibitemOpen
  \bibfield  {author} {\bibinfo {author} {\bibfnamefont {H.-D.}\ \bibnamefont
  {Meyer}}, \bibinfo {author} {\bibfnamefont {F.}~\bibnamefont {{Le
  Qu\'er\'e}}}, \bibinfo {author} {\bibfnamefont {C.}~\bibnamefont
  {L\'eonard}}, \ and\ \bibinfo {author} {\bibfnamefont {F.}~\bibnamefont
  {Gatti}},\ }\href@noop {} {\bibfield  {journal} {\bibinfo  {journal} {Chem.\
  Phys.}\ }\textbf {\bibinfo {volume} {329}},\ \bibinfo {pages} {179} (\bibinfo
  {year} {2006})}\BibitemShut {NoStop}%
\bibitem [{\citenamefont {Doriol}\ \emph {et~al.}(2008)\citenamefont {Doriol},
  \citenamefont {Gatti}, \citenamefont {Iung},\ and\ \citenamefont
  {Meyer}}]{dor08:224109}%
  \BibitemOpen
  \bibfield  {author} {\bibinfo {author} {\bibfnamefont {L.~J.}\ \bibnamefont
  {Doriol}}, \bibinfo {author} {\bibfnamefont {F.}~\bibnamefont {Gatti}},
  \bibinfo {author} {\bibfnamefont {C.}~\bibnamefont {Iung}}, \ and\ \bibinfo
  {author} {\bibfnamefont {H.-D.}\ \bibnamefont {Meyer}},\ }\href@noop {}
  {\bibfield  {journal} {\bibinfo  {journal} {J.~Chem.\ Phys.}\ }\textbf
  {\bibinfo {volume} {129}},\ \bibinfo {pages} {224109} (\bibinfo {year}
  {2008})}\BibitemShut {NoStop}%
\bibitem [{\citenamefont {Meyer}(2012)}]{mey12:351}%
  \BibitemOpen
  \bibfield  {author} {\bibinfo {author} {\bibfnamefont {H.-D.}\ \bibnamefont
  {Meyer}},\ }\href {\doibase 10.1002/wcms.87} {\bibfield  {journal} {\bibinfo
  {journal} {Wiley Interdiscip. Rev.: Comp. Mol. Sci.}\ }\textbf {\bibinfo
  {volume} {2}},\ \bibinfo {pages} {351} (\bibinfo {year} {2012})}\BibitemShut
  {NoStop}%
\bibitem [{\citenamefont {Worth}\ \emph {et~al.}()\citenamefont {Worth},
  \citenamefont {Beck}, \citenamefont {J{\"a}ckle}, \citenamefont {Vendrell},\
  and\ \citenamefont {Meyer}}]{mctdh:MLpackage}%
  \BibitemOpen
  \bibfield  {author} {\bibinfo {author} {\bibfnamefont {G.~A.}\ \bibnamefont
  {Worth}}, \bibinfo {author} {\bibfnamefont {M.~H.}\ \bibnamefont {Beck}},
  \bibinfo {author} {\bibfnamefont {A.}~\bibnamefont {J{\"a}ckle}}, \bibinfo
  {author} {\bibfnamefont {O.}~\bibnamefont {Vendrell}}, \ and\ \bibinfo
  {author} {\bibfnamefont {H.-D.}\ \bibnamefont {Meyer}},\ }\href@noop {}
  {}\bibinfo {howpublished} {{The MCTDH Package, Version 8.2, (2000). H.-D.
  Meyer, Version 8.3 (2002), {V}ersion 8.4 (2007). O. Vendrell and H.-D. Meyer
  {V}ersion 8.5 (2013). Versions 8.5 and 8.6 contains the ML-MCTDH algorithm.
  Used version: 8.6.1 (May 2022). {S}ee http://mctdh.uni-hd.de/}}\BibitemShut
  {NoStop}%
\bibitem [{\citenamefont {Harris}\ \emph {et~al.}(1965)\citenamefont {Harris},
  \citenamefont {Engerholm},\ and\ \citenamefont
  {Gwinn}}]{Harris1965Calculation}%
  \BibitemOpen
  \bibfield  {author} {\bibinfo {author} {\bibfnamefont {D.~O.}\ \bibnamefont
  {Harris}}, \bibinfo {author} {\bibfnamefont {G.~G.}\ \bibnamefont
  {Engerholm}}, \ and\ \bibinfo {author} {\bibfnamefont {W.~D.}\ \bibnamefont
  {Gwinn}},\ }\href {\doibase 10.1063/1.1696963} {\bibfield  {journal}
  {\bibinfo  {journal} {J. Chem. Phys.}\ }\textbf {\bibinfo {volume} {43}},\
  \bibinfo {pages} {1515} (\bibinfo {year} {1965})}\BibitemShut {NoStop}%
\bibitem [{\citenamefont {Hummel}\ \emph
  {et~al.}(2021{\natexlab{b}})\citenamefont {Hummel}, \citenamefont {Eiles},\
  and\ \citenamefont {Schmelcher}}]{Hummel2021Synthetic}%
  \BibitemOpen
  \bibfield  {author} {\bibinfo {author} {\bibfnamefont {F.}~\bibnamefont
  {Hummel}}, \bibinfo {author} {\bibfnamefont {M.~T.}\ \bibnamefont {Eiles}}, \
  and\ \bibinfo {author} {\bibfnamefont {P.}~\bibnamefont {Schmelcher}},\
  }\href {\doibase 10.1103/PhysRevLett.127.023003} {\bibfield  {journal}
  {\bibinfo  {journal} {Phys. Rev. Lett.}\ }\textbf {\bibinfo {volume} {127}},\
  \bibinfo {pages} {023003} (\bibinfo {year} {2021}{\natexlab{b}})}\BibitemShut
  {NoStop}%
\end{thebibliography}%
\end{document}